\documentclass{article}%
\usepackage{amsmath}
\usepackage{amsfonts}
\usepackage{amssymb}
\usepackage{graphicx}%
\newcommand{\bpartial}{\mathop{\partial\kern -4pt\raisebox{.8pt}{$|$}}}
\newcommand{\bra}{\mathopen{[\kern-1.6pt[}}
\newcommand{\ket}{\mathclose{]\kern-1.5pt]}}
\newcommand{\bbra}{\mathopen{[\kern-2.2pt[\kern-2.3pt[}}
\newcommand{\bket}{\mathclose{]\kern-2.1pt]\kern-2.3pt]}}

\makeindex
\begin{document}

\title {\large{\bf CLASSIFICATION OF FOUR DIMENSIONAL REAL LIE BIALGEBRAS OF SYMPLECTIC TYPE AND THEIR POISSON-LIE GROUPS}}
\vspace{3mm}
\author { \small{ \bf J. Abedi-Fardad$^1$ }\hspace{-2mm}{ \footnote{ e-mail: j.abedifardad@bonabu.ac.ir}}, \small{ \bf A.
Rezaei-Aghdam$^2$ }\hspace{-1mm}{\footnote{ e-mail: rezaei-a@azaruniv.edu - Corresponding author.}}, \small{ \bf  Gh. Haghighatdoost$^3$}\hspace{-1mm}{ \footnote{ e-mail: gorbanali@azaruniv.edu}}  \\
{\small{$^{1,3}$\em Department of Mathematics,
University of Bonab , Tabriz, Iran}}\\
{\small{$^{2}$\em Department of Physics, Azarbaijan Shahid Madani
University, 53714-161, Tabriz, Iran}}\\
 {\small{$^{3}$\em Department
of Mathematics,Azarbaijan Shahid Madani University, 53714-161, Tabriz, Iran}} }
 \maketitle

\begin{abstract}
 In this paper we classify all  four dimensional real   Lie bialgebras of symplectic type.
  The classical r-matrices for these Lie bialgebras and  Poisson structures on  all of the related four dimensional
Poisson-Lie groups are also obtained. Some new
integrable models for which  the Poisson-Lie group plays the role as a phase space and its dual Lie group plays the role of a symmetry group of the system, are obtained.
\end{abstract}

\smallskip

{\bf keywords:}{  Lie bialgebra, Poisson-Lie group,
Classical r-matrix, Integrable systems.}

\section {\large {\bf Introduction}}

The theory of classical integrable systems is formally related to
the geometry and representation theory  of Poisson-Lie groups,
their Lie bialgebras \cite{Drin}, \cite{VGD} and the corresponding classical
r-matrices  \cite{MAS} (for a review see  \cite{YKS}). There is a
detailed classification of r-matrices only for the complex
semi-simple Lie algebras  \cite{AAB}. Recently, non-semisimple Lie
algebras has shown to play important role in  physical problems. Of course
there are attempts for the classification of low dimensional
non-semisimple Lie bialgebras
 [\cite{JMF} - \cite{RHR}] and  Lie superbialgebras  \cite{ER1}. In this paper we will try to classify four dimensional
 real  Lie bialgebras of symplectic type such that on the Lie algebras ${\bf g} $ and their duals ${\tilde {\bf g}} $ we have symplectic structures. We will also obtain the classical r-matrices and Poisson structures on the corresponding Poisson-Lie groups. The reason for obtaining this classification is that
 we are interested in constructing physical models in which the Lie group $G$ (of the Lie algebra ${\bf g} $) plays the role of phase space and
the dual Lie group ${\tilde{G}}$ (of the dual Lie algebra ${\tilde {\bf g}} $ ) plays the role of symmetry group of the systems (or vice versa).
 The outline of the paper is as follows. In section  two, we briefly review the definitions and notations.
  In section three, after giving the list of four dimensional real  Lie algebra of symplectic type \cite{GO}, \cite{MS}  based in to \cite{JP}
 (classification of real four dimensional Lie algebras); we classify four dimensional real  Lie bialgebras of symplectic type according to the method given in \cite{ER1}. In section four, we determine the coboundary Lie bialgebras from the list  obtained in section three. Then, in section five, using Sklyanin  bracket, we calculate the Poisson structures on the Poisson-Lie groups. Also for the noncoboundary Lie bialgebras using
  adjoint matrices we perform the same calculations. The complete lists of compatible symplectic structures on real four dimensional Poisson-Lie groups
  are given in tables 8 and 9. Finally in section six, two integrable systems as examples of physical applications are obtained. For one of
   these examples,
  the Lie group $\bf{A_{4,9}^{-\frac{1}{2}}}$  plays the role of phase space and the dual Lie group $\bf{A_{4,9}^1.ii}$  plays the role of symmetry group of the systems and in the second example these roles are exchanged. Some conclusion
  remarks are given in section seven. Other computational results are given in appendices A-D.


\section {\large {\bf Definitions and notations }}

In this section, we  recall some basic definitions and propositions
of Lie bialgebras  \cite{Drin},\cite{VGD} (see for a review
\cite{YKS} ) . Let ${\bf g} $ be a finite dimensional Lie algebra
and  ${\bf g}^\ast$ be its dual space with  respect to a
non-degenerate canonical pairing ( . , . ) .

\goodbreak

{\bf Definition}: A {\em Lie bialgebra} is a Lie algebra ${\bf g}$ with a skew-symmetric linear map
$\delta : \bf g \rightarrow  \bf g\otimes \bf g$ such that:\\
a) $\delta$ is a one-cocycle, i.e.:
\begin{equation}
\delta ([X,Y])=[\delta (X),1\otimes Y+Y\otimes 1]+[1\otimes X+X\otimes 1,\delta(Y)]\;\;\;\;\;\;\;\;\forall X,Y\in \bf g,
\end{equation}
where 1 is the identity map on $\bf g$.\\
 b) ${\delta}^t :{\bf
g}^\ast \otimes {\bf g}^\ast \rightarrow {\bf g}^\ast$ is a Lie
bracket on ${\bf g}^\ast$:
\begin{equation}\label{BA6}
(\xi\otimes \eta ,\delta(X))=({\delta}^t(\xi\otimes \eta),X)=([\xi ,\eta]_{{\bf g}^\ast} ,X)\;\;\;\;\;\;\;\;\forall X\in \bf g;\;\; \xi,\eta\in {\bf g}^\ast.
\end{equation}
 The Lie bialgebra defined in this way will be denoted by $(\bf g,{\bf g}^\ast)$ or $(\bf g ,\delta)$ .

{\bf Proposition}:\cite{RHR} If there exists an automorphism $A$ of
$\bf g$, such that
 \begin{equation}\label{BB9}
 \delta^\prime = (A\otimes A)\circ \delta \circ A^{-1},
 \end{equation}
then the one-cocycles $\delta$ and $\delta^\prime$ of the Lie algebra $\bf g$ are equivalent. In this case the
two Lie bialgebras $(\bf g,\delta)$ and $(\bf g,\delta^\prime)$ are equivalent.

{\bf Definition}:  Lie bialgebra  $(\bf g,{\bf g}^\ast)$  is  called a {\em coboundary} Lie bialgebra if there exists an element
 $r\in \bf g\otimes\bf g$ such that:
\begin{equation}\label{BA1}
\delta (X)=[1\otimes X+X\otimes 1,r ]\;\;\;\;\;\;\forall X\in \bf g.
\end{equation}

{\bf Definition}: Coboundary Lie bialgebras can be one  of the  two  following  different types:\\
{\bf a)} If $r$ is a skew-symmetric solution of the classical Yang-Baxter equation (CYBE):
\begin{equation}
[[r,r]]=0,
\end{equation}
then the coboundary Lie bialgebra is said to be triangular; where
the Schouten bracket is defined by:
\begin{equation}
[[r,r]]=[r_{12},r_{13}]+[r_{12},r_{23}]+[r_{13},r_{23}],
\end{equation}
such that for $r=r^{ij}X_i\otimes X_j$, we have
$r_{12}=r^{ij}X_i\otimes X_j\otimes 1$, $r_{13}=r^{ij}X_i\otimes 1
\otimes X_j$ and  $r_{23}=r^{ij} 1\otimes X_i\otimes X_j$, in which
$\lbrace X_i \rbrace$ is the basis of the Lie algebra ${\bf g}$.\\
{\bf b)} If $r$ is a solution of CYBE, such that $r_{12}+r_{21} $ is
a $\bf g$ invariant element of $\bf g \otimes \bf g$; then, the
coboundary  Lie bialgebra is said to be quasi-triangular. Sometimes, the
condition b) can be replaced with the following one  \cite{VGD},
\cite{VC}:\\
 {\bf $\bf b^\prime$)} If $r$ is a skew-symmetric solution
of the modified CYBE:
\begin{equation}
[[r,r]]=\omega \;\;\;\;\;\;\;\;\;\;\;\omega \in \wedge^{3}{\bf g},
\end{equation}
 then coboundary Lie bialgebra is said to be quasi-triangular.
We note that if $\bf g$ is a Lie bialgebra then ${\bf g}^\ast$ is
also a Lie bialgebra  \cite{VC}, while, this is not always true for the
coboundary property.\\

{\bf Definition}:   {\it Manin triple}   is a triple of Lie algebras
$(\cal{D} , {\bf g} , {\bf \tilde{g}})$  with a non-degenerate
ad-invariant symmetric bilinear form $<.~ , ~. >$ on $\cal{D}$, so
that

 \hspace{2mm} 1.~~${\bf g}$ and ${\bf \tilde{g}}$ are Lie
subalgebras of $\cal{D}$,

 \hspace{2mm} 2.~~$\cal{D} = {\bf
g}\oplus{\bf \tilde{g}}$ as a vector space,

 \hspace{2mm}
 3.~~${\bf g}$ and ${\bf
\tilde{g}}$ are isotropic with respect to $< .~, ~.
>$,  i.e.,
\begin{equation}\label{BB5}
<X_i , X_j> = <\tilde{X}^i , \tilde{X}^j> = 0, \hspace{10mm}
\;<X_i , \tilde{X}^j>\; ={\delta^j}_i.
\end{equation}

There is a one-to-one correspondence between  Manin triple $(\cal{D} , {\bf
g} , {\bf \tilde{g}})$ with ${\bf \tilde{g}}= {\bf g}^\ast$ and  Lie
bialgebra $({\bf g},{\bf g}^\ast)$ \cite{VC}. If we choose the
structure constants of Lie algebras ${\bf g}$ and ${\bf \tilde{g}}$
as follows:
\begin{equation}\label{BA2}
[X_i , X_j] = f_{ij}^{\;\;\;k} X_k,\hspace{20mm} [\tilde{X}^i ,\tilde{ X}^j] ={{\tilde{f}}^{ij}}_{\; \; \: k} {\tilde{X}^k}, \\
\end{equation}
with them satisfying
the following Jacobi identities;
\begin{equation}\label{BA7}
f_{ij}^{\;\;\;k} f_{km}^{\;\;\;n}+f_{ik}^{\;\;\;n} f_{mj}^{\;\;\;k}+f_{jk}^{\;\;\;n} f_{im}^{\;\;\;k}=0 ,
\end{equation}
\begin{equation}\label{BA8}
{\tilde{f}}^{ij}_{\;\; \;  k} \; {\tilde{f}}^{km}_{\;\; \; \; n} +
{\tilde{f}}^{im}_{\;\; \; \; k} \; {\tilde{f}}^{jk}_{\;\; \; \; n}
+{\tilde{f}}^{jm}_{\;\; \; \; k} \; {\tilde{f}}^{ki}_{\;\; \; \;
n}=0 ,
\end{equation}
then, ad-invariance of the bilinear form $< .~ , ~. >$ on $\cal{D} =
{\bf g}\oplus{\bf \tilde{g}}$ implies that \cite { VC}-\cite {YKS}
\begin{equation}
[X_i , \tilde{X}^j] ={\tilde{f}^{jk}}_{\; \; \; \:i} X_k +
f_{ki}^{\;\;\;j} \tilde{X}^k,
\end{equation}
where, using  (\ref{BB5}) ,(\ref{BA2})  and (\ref{BA6}) we obtain
\begin{equation}\label{BA3}
\delta(X_i) = {\tilde{f}^{jk}}_{\; \; \; \:i} X_j \otimes
X_k.
\end{equation}
 By applying the above relation in (1), one can obtain the following  mixed Jacobi relation:
\begin{equation}\label{BB1}
f_{kl}^{\;\;\;m}{\tilde{f}^{ij}}_{\; \; \; \; m}=
f_{mk}^{\;\;\;i}{\tilde{f}^{jm}}_{\; \; \; \; \; l} -
f_{ml}^{\;\;\;i}{\tilde{f}^{jm}}_{\; \; \; \; \; k}-
f_{mk}^{\;\;\;j}{\tilde{f}^{im}}_{\; \; \; \; \; l}+
f_{ml}^{\;\;\;j}{\tilde{f}^{im}}_{\; \; \; \; \; k}.
\end{equation}
This relation can also be obtained from  Jacobi identity of
$\cal{D}$.
 \vspace{3mm}

\section {\large {\bf Classification of four dimensional real  Lie bialgebras of symplectic type. }}

In this section, we  classify four dimensional real  Lie
bialgebras of symplectic type, such that  we have symplectic structures on the Lie
algebras ${\bf g}$ and their duals ${\tilde{\bf g}}$. The
four dimensional real Lie algebras of symplectic type have been  classified in
\cite{GO}. Here, we will use the same method which previously have
been  considered  in the classification of  Lie superbialgebras  in
\cite{ER1}. Let us first have a short review about symplectic
structures on a Lie algebras and four dimensional real
Lie algebras of symplectic type and then have a short review about classification
method of low dimensional real Lie bialgebras.

\subsection {\large {\bf Four dimensional real   Lie algebras of symplectic type }}
A symplectic structure $\omega$ on a   2n-dimensional Lie algebra ${\bf g}$  is defined as  a two form such that
 \footnote{Here we use the cohomology of a Lie algebra such that $d$ is a extrinsic
 derivative on the Lie algebra ${\bf g}$ (see for example \cite{DAA}).}\hspace{2mm}

1)  $\omega$ is closed , i.e., $ d\omega =0$;\hspace{2mm}

2)  $\omega$ has maximal rank, that is,  $\omega ^n$ is a volume form on the corresponding Lie group.
 The list of four dimensional real Lie algebras with symplectic  structure is given in \cite{GO} (see also \cite{MS} );
  and we brought it in table 1 for self containing of  the paper
  \footnote{Note that in table 1 we use the Patera classification \cite{JP}  of  four dimensional real Lie  algebras.}
\\

{\small {\bf Table 1}}: {\small
Four dimensional  real  Lie algebras of symplectic type.}\\
    \begin{tabular}{l l | l l   }
    \hline\hline
{\footnotesize ${\bf g}$ }&{\footnotesize Non-zero commutation relations }&{\footnotesize ${\bf g}$ }&{\footnotesize Non-zero commutation relations } \\ \hline

{\footnotesize$A_{4,1}$}&{\footnotesize$[X_2,X_4]=X_1 \;,\;\;[X_3,X_4]=X_2$} &{\footnotesize$A_{4,2}^{-1}$}&{\footnotesize$[X_1,X_4]=-X_1 ,\;[X_2,X_4]=X_2 ,\;[X_3,X_4]=X_2+X_3$} \\

{\footnotesize$A_{4,3}$}&{\footnotesize$[X_1,X_4]=X_1 \;,\;\;[X_3,X_4]=X_2$} &{\footnotesize$A_{4,5}^{-1,-1}$}&{\footnotesize$[X_1,X_4]=X_1 \;,\;\;[X_2,X_4]=-X_2 \;,\;\;[X_3,X_4]=-X_3$} \\

{\footnotesize$A_{4,5}^{-1,b}$}&{\footnotesize$[X_1,X_4]=X_1 , [X_2,X_4]=-X_2 , [X_3,X_4]=b \;X_3$} &{\footnotesize$A_{4,5}^{a,-1}$}&{\footnotesize$[X_1,X_4]=X_1 \;,\;\;[X_2,X_4]=a\;X_2 \;,\;\;[X_3,X_4]=-X_3$} \\

{\footnotesize$A_{4,5}^{a,-a}$}&{\footnotesize$[X_1,X_4]=X_1 , [X_2,X_4]=a X_2 , [X_3,X_4]=-a X_3$} &{\footnotesize$A_{4,6}^{a,0}$}&{\footnotesize$[X_1,X_4]=a X_1 \;,\;\;[X_2,X_4]=-X_3 \;,\;\;[X_3,X_4]=X_2$} \\

{\footnotesize$A_{4,7}$}&{\footnotesize$[X_1,X_4]=2X_1 ,\;[X_2,X_3]=X_1 ,\;[X_2,X_4]=X_2$} &{\footnotesize$A_{4,9}^{0}$}&{\footnotesize$[X_1,X_4]=X_1 \;,\;\;[X_2,X_3]=X_1 \;,\;\;[X_2,X_4]=X_2$} \\

&{\footnotesize$\;[X_3,X_4]=X_2+X_3$} &{\footnotesize$A_{4,9}^{-1/2}$}&{\footnotesize$[X_1,X_4]=1/2 X_1 \;,\;\;[X_2,X_3]=X_1 \;,\;\;[X_2,X_4]=X_2$} \\

{\footnotesize$A_{4,9}^1$}&{\footnotesize$[X_1,X_4]=2X_1 ,\;[X_2,X_3]=X_1 ,\;[X_2,X_4]=X_2$} &&{\footnotesize$[X_3,X_4]=-1/2X_3 $} \\

&{\footnotesize$\;[X_3,X_4]=X_3$} &{\footnotesize$A_{4,9}^{b}$}&{\footnotesize$[X_1,X_4]=(1+b) X_1 ,\;[X_2,X_3]=X_1 ,\;[X_2,X_4]=X_2$} \\

{\footnotesize$A_{4,11}^b$}&{\footnotesize$[X_1,X_4]=2a X_1 \;,\;\;[X_2,X_3]=X_1 $} &&{\footnotesize$[X_3,X_4]=b X_3 $} \\

&{\footnotesize$\;[X_2,X_4]=a X_2-X_3\;,\;\;[X_3,X_4]=X_2+aX_3$} &{\footnotesize$A_{4,12}$}&{\footnotesize$[X_1,X_3]= X_1 \;,\;\;[X_1,X_4]=-X_2 \;,\;\;[X_2,X_3]=X_2$} \\

{\footnotesize$A_2\oplus A_2$}&{\footnotesize$[X_1,X_2]= X_2 \;,\;\;[X_3,X_4]=X_4 $} &&{\footnotesize$[X_2,X_4]= X_1 $} \\

{\footnotesize$VI_0\oplus R$}&{\footnotesize$[X_1,X_3]=X_2 \;,\;\;[X_2,X_3]=X_1 $} &{\footnotesize$III\oplus R$}&{\footnotesize$[X_1,X_2]=-X_2-X_3 \;,\;\;[X_1,X_3]=-X_2-X_3 $} \\

{\footnotesize$VII_0\oplus R$}&{\footnotesize$[X_1,X_3]=-X_2 \;,\;\;[X_2,X_3]=X_1 $} &{\footnotesize$II\oplus R$}&{\footnotesize$[X_2,X_3]=X_1 $} \\

\smallskip \\
\hline\hline
\end{tabular}

\subsection {\large {\bf Review of the  classification method for low dimensional real  Lie bialgebras }}
In this section, we review the method of obtaining and classification
of low dimensional real Lie bialgebras, as applied  for first time
in \cite{ER1} for classification of real  low dimensional  Lie
superbialgebras. For this proposes, we use the following adjoint
representation:
$$ ({\cal X}_i)_j^{\;k}=-{f_{ij}}^k ,\;\;\;\;\;\;({\cal Y}^k)_{ij}=-{f_{ij}}^k , $$
\begin{equation}
(\tilde{\cal X}^i)^j_{\;k}=-{\tilde{f}^{ij}}_{\;\;\;k },\;\;\;\;\;\;(\tilde{\cal  Y}_k)^{ij}=-{\tilde{f}^{ij}}_{\;\;\;k},
\end{equation}
for writing the matrix forms of equations   (\ref{BA8}) and (\ref{BB1})  as follows, respectively,
\begin{equation}\label{BB2}
(\tilde{\cal X}^i)^j_{\;k} \tilde{\cal X}^k +\tilde{\cal X}^i \tilde{\cal X}^j - \tilde{\cal X}^j \tilde{\cal X}^i=0 ,
\end{equation}
\begin{equation}\label{BB3}
 (\tilde{\cal X}^i)^j_{\;l} {\cal Y}^l= - (\tilde{{\cal X}}^t)^j {\cal Y}^i+{\cal Y}^j \tilde{{\cal X}}^i-{\cal Y}^i \tilde{{\cal X}}^j+(\tilde{{\cal X}}^t)^i {\cal Y}^j .
\end{equation}
Having the structure constants of the Lie algebra ${\bf g}$
$({f_{ij}}^{\;k})$ , we solve matrix equations  (\ref{BB2}) and
(\ref{BB3}) in order to obtain the structure constants of the dual
Lie algebras $\tilde{\bf g}$ (${\tilde{f}^{ij}}_{\;\;\;k }$), such
that $(\bf g,\tilde{\bf g})$ is a  Lie bialgebra. By this method we
will classify four dimensional real Lie bialgebras of symplectic type.
 We will perform this task in the following three steps similar to \cite{ER1}.\\

{\bf Step1: }Solving  equations (\ref{BB2}) and (\ref{BB3}) and determining the Lie algebras $\bf g^\prime$
which are isomorphic with dual solutions $\tilde{\bf g}$.

With the solution of matrix equations (\ref{BB2}) and (\ref{BB3})
for obtaining matrices $\tilde{{\cal X}}^i$, some structure
constants of $\tilde{\bf g}$ are obtained to be zero, some unknown
and some obtained in terms of each other. In order to know whether
$\tilde{\bf g}$ is one of the Lie algebras of table or isomorphic to
them, we must use the following isomorphic relation between the
obtained Lie algebras $\tilde{\bf g}$ and one of the Lie algebras of
table 1, e.g., ${\bf g}^\prime$ .Applying the following
transformation for a change of basis $\tilde{\bf g}$ as follows
\begin{equation}
{\tilde{X}^{\prime \; i}}=C_{\;j}^i  {\tilde{X}^{
j}},\;\;\;\;\;\;\;[{\tilde{X}^{\prime \; i}},{\tilde{X}^{\prime
\;j}}]= {\tilde{f}^{\prime ij} }_{\;\;\;\;k } {\tilde{X}^{\prime \;
k}},
\end{equation}
 we have the following matrix equations for isomorphism:
\begin{equation}\label{BB4}
C(C_{\;j}^i  {\tilde{X}^{ j}}_{(\tilde{\bf g})})={\tilde{X}^{ i}}_{({\bf g}^\prime)} C .
\end{equation}
Solving  (\ref{BB4}) with the condition $det\; C\neq 0$ , we obtain
some extra conditions on ${\tilde{f}^{kl}}_{(\tilde{\bf g})\;m}$'s
as imposed by (\ref{BB2}) and (\ref{BB3}).

{\bf Step2: }Calculate the general form of the transformation matrices
$B:{\bf g}^\prime\rightarrow {\bf g}^\prime .i$; such that $({\bf
g},{\bf g^\prime}.i)$ are Lie bialgebras.

 As the second step we
transform Lie bialgebra  $(\bf g,\tilde{\bf g})$  to Lie  bialgebra
$({\bf g},{\bf g^\prime}.i)$(where ${\bf g^\prime}.i$ is isomorphic
as Lie algebra   ${\bf g^\prime}$ ) with an automorphism of the Lie
algebra $\bf g$. As inner product (\ref{BB5}) is invariant, we have
$A^{-t}:\tilde{\bf g}\rightarrow {\bf g}^\prime .i$
\begin{equation}
{{X^\prime}_{ i}}=A_{i}^{\;k} {X}_{k },\;\;\;\;\; {\tilde{X}^{\prime
j}}=(A^{-t})_{\;l}^{j} \tilde{X}^{l},\;\;\;\;\;<X_i^\prime ,
\tilde{X}^{\prime j}>\; ={\delta^j}_i,
\end{equation}
where $A^{-t}$ is the inverse transpose of   same matrix  $A\in Aut({\bf g})$. Thus, we have the following relation
\begin{equation}\label{BB6}
(A^{-t})^i_{\;k} \tilde{f}^{kl}_{(\tilde{\bf g})\;m} (A^{-t})^j_{\;l}={f}^{ij}_{({\bf g}^\prime .i)\;n} (A^{-t})^n_{\;m}.
\end{equation}
Now, for derivation of Lie bialgebras $({\bf g},{\bf g^\prime}.i)$,  we must find the Lie algebras ${\bf g^\prime}.i$ or transformations
 $B:{\bf g}^\prime\rightarrow {\bf g}^\prime .i$,  such that
\begin{equation}\label{BB7}
B^i_{\;k} {f}^{kl}_{({\bf g}^\prime )\;m} B^j_{\;l}={f}^{ij}_{({\bf g}^\prime .i)\;n}  B^n_{\;m} .
\end{equation}
For this purpose, it is enough to eliminate ${f}^{ij}_{({\bf g}^\prime .i)\;n}$ between (\ref{BB6}) and (\ref{BB7}). Then, we  have the
following matrix equation for $B$:
\begin{equation}\label{BB8}
(A^{-t})^i_{\;m} \tilde{\cal X}^{t\;m}_{\tilde{(\bf g)}} A^{-1}=(B^t A)^{-1} (B^i_{\;k}{ \cal X}^{t\;k}_{({\bf g}^\prime)}) B^t .
\end{equation}
Now, by solving  (\ref{BB8}), we obtain the general form of the matrix $B$
with the condition $det\;B\neq 0$.

{\bf Step 3: }Calculate and classify the non-equivalent Lie
bialgebras.

 Having solved (\ref{BB8}) , we obtain the general form
of the matrix $B$ so that its elements are written in terms of the
elements of matrices $A, C$ and structure constants
$\tilde{f}^{ij}_{(\tilde{\bf g})\;k}$.
 Now after substituting $ B$ in (\ref{BB7}), we obtain the structure constants ${f}^{ij}_{({\bf g^\prime}.i)\;n}$ of the Lie algebra
${\bf g^\prime} .i$ in terms of elements of matrices $A$ and $C$ and
structure constants $\tilde{f}^{ij}_{(\tilde{\bf g})\;k}$. Then, it is checked whether it
is possible to equalize the structure constants ${f}^{ij}_{({\bf
g^\prime}.i)\;n}$ with each other and with $\pm1 $ or not, such that $det \;B\neq 0 , det \;A\neq 0 $ and $det \;C\neq 0$. In this
way, we obtain matrices  $B_1, B_2,$ etc. Note that in obtaining $B_i
s$ we impose the condition $B B_i^{-1} \in Aut ^t( \bf g)$.
 If this condition is not satisfied then we cannot impose it on the structure constants, because $B$ and $B_i$ are not equivalent (see below).\\
Now, using isomorphism of matrices $B_1 , B_2$, etc., we can obtain Lie
bialgebras $({\bf g},{\bf g}^\prime .i)$, $({\bf g},{\bf g}^\prime
.ii)$, etc. On the other hand, there remains a question of whether these
Lie bialgebras are equivalent? In order to answer this question, we
use the matrix form of relation (\ref{BB9}). Consider the two Lie
bialgebras $({\bf g},{\bf g}^\prime .i)$, $({\bf g},{\bf g}^\prime
.ii)$; then using
\begin{equation}
A(X_i)=A_i^{\;j}X_j,
\end{equation}
the relation  (\ref{BB9}) will have the following matrix form:
\begin{equation}\label{BB10}
A^t((A^t)^i_{\;k} {\cal X}_{({\bf g}^\prime .i)}^{\;\;k})={\cal X}_{({\bf g}^\prime .ii)}^{\;\;i} A^t.
\end{equation}
 On the other hand, the transformation matrix between ${\bf g}^\prime .i$ and ${\bf g}^\prime .ii$ is  $B_2 B_1^{-1}$  such that
$B_1:{\bf g}^\prime\rightarrow {\bf g}^\prime .i$  and  $B_2:{\bf g}^\prime\rightarrow {\bf g}^\prime .ii$, then we have
\begin{equation}\label{BB11}
(B_2 B_1^{-1} ) ((B_2 B_1^{-1})^i_{\;k}{ \cal X}_{({\bf g}^\prime .i)}^{\;\;k})={\cal X}_{({\bf g}^\prime .ii)}^{\;\;i}(B_2 B_1^{-1}).
\end{equation}
A comparison of  (\ref{BB11}) with  (\ref{BB10}) reveals that if
$B_2 B_1^{-1}\in A^t$ , then, the Lie bialgebras $({\bf g},{\bf
g}^\prime .i)$ and $({\bf g},{\bf g}^\prime .ii)$ are equivalent. In
this way, we obtain non-equivalent class of $B_i s$, and we consider
only one element of this class. Now, we will use this method
for  obtaining  and classifying all four dimensional real
Lie bialgebras of symplectic type.

\subsection {\large {\bf Classification of four dimensional real  Lie bialgebra of symplectic type }}

In the following  we explain the above
method   by describing the details of the calculations for obtaining
the symplectic Lie bialgebra $(A_{4,1},(II\oplus R).i)$.

\smallskip
{\bf { An example: }}

One of the solutions of Jacobi and mixed Jacobi identities  (\ref{BB2}) and (\ref{BB3}) for the Lie algebra  $A_{4,1}$ has the following form:
\begin{equation}
{\tilde{f}^{12}}_{\;\;\;2 }={\tilde{f}^{14}}_{\;\;\;4 }=\alpha , \; {\tilde{f}^{12}}_{\;\;\;3 }=\beta ,\;{\tilde{f}^{12}}_{\;\;\;4}=\gamma ,\; {\tilde{f}^{23}}_{\;\;\;4 }=\lambda ,
\end{equation}
where $ \alpha , \beta , \gamma , \lambda $ are arbitrary constants. Now, using (\ref{BB4}) the corresponding Lie algebra ${\tilde{\bf g}} $ is isomorphic to the Lie algebra $II\oplus R$, with the
following isomorphism matrix
\begin{equation}
C=\left(
\begin{matrix}
c_{11}&c_{12}&c_{13}&c_{14}  \cr
 c_{21}&c_{22}&c_{23}&c_{24} \cr
 \frac{-c_{13} c_{22}+c_{12} c_{23}-c_{41} \gamma}{\beta}&0&0&c_{34}\cr
 c_{41}&0&0&-\frac{c_{34 }\beta}{\gamma}
 \end{matrix}
\right),
\end{equation}
satisfying the conditions $ \alpha =0$ and $ \lambda =0$, where,  $c_{ij} \in \Re $ are arbitrary elements of the isomorphism matrices.
 Now, by substituting the above results and using  the following  automorphism group element of
the Lie algebra  $A_{4,1}$, we have

\begin{equation}
A=\left(
\begin{matrix}
a_{22} a_{44}&0&0&0 \cr
 a_{32}a_{44}&a_{22}&0&0 \cr
a_{31}&a_{32}&\frac{a_{22}}{a_{44}}&0\cr
a_{41}&a_{42}&a_{43}& a_{44}

 \end{matrix}
\right),
\end{equation}
 where $a_{22} a_{44}\neq 0$ and $a_{ij}$ are arbitrary elements of the $A$
 matrices. In (\ref{BB8}), one can obtain the following form for the matrix $B$

\begin{equation}
B=\left(
\begin{matrix}
b_{11}&b_{12}&b_{13}&b_{14}  \cr
 b_{21}&b_{22}&b_{23}&b_{24} \cr
 \frac{- a_{43} a_{44} b_{41} \beta- a_{22}^2 a_{44}^2(b_{13} b_{22}-b_{12} b_{23})-a_{44}^2 b_{41} \gamma }{a_{22} \beta}&0&0&-\frac{a_{43} a_{44} b_{44}\beta+ a_{44}^2 b_{44} \gamma}{ a_{22} \beta}\cr
 b_{41}&0&0& b_{44}

 \end{matrix}
\right),
\end{equation}
where $b_{ij}$ are arbitrary elements of the $B$  matrices. Now,
using (\ref{BB7}) we have found the following commutation relations
for the algebra ${\bf g ^\prime} .i$
\begin{equation}
[{\tilde X}^1,{\tilde X}^2]= \frac{\beta}{a_{22} \;a_{44}^2}{\tilde X}^3+\frac{a_{43}\; \beta + a_{44}\; \gamma }{a_{22}^2\; a_{44}}{\tilde X}^4 .
\end{equation}
On the other hand,  by choosing $a_{43}=0$, we have
 \begin{equation}
B_1=\left(
\begin{matrix}
b_{11}&b_{12}&b_{13}&b_{14}  \cr
 b_{21}&b_{22}&b_{23}&b_{24} \cr
 \frac{- a_{22}^2 a_{44}^2(b_{13} b_{22}-b_{12} b_{23})-a_{44}^2 b_{41} \gamma }{a_{22} \beta}&0&0& -\frac{ a_{44}^2 b_{44} \gamma}{ a_{22} \beta}\cr
 b_{41}&0&0& b_{44}
 \end{matrix}
\right),
\end{equation}
where, since $B_1 B^{-1} \in A^t  $, then, $ B_1$  is equivalent to $ B $.
Furthermore,  by choosing $\frac{\beta}{a_{22} \;a_{44}^2} =\frac{
\gamma }{a_{22}^2}=q\neq0 $, we have

  \begin{equation}
B_2=\left(
\begin{matrix}
b_{11}&b_{12}&b_{13}&b_{14}  \cr
 b_{21}&b_{22}&b_{23}&b_{24} \cr
 \frac{- (b_{13} b_{22}-b_{12} b_{23}) }{q} - b_{41} &0&0&-b_{44} \cr
 b_{41}&0&0& b_{44}

 \end{matrix}
\right),
\end{equation}
where, since $B_2 B_1^{-1} \in A^t  $, then, $ B_2$  is equivalent to $ B _1$.
 Now, by choosing $q=1$, we have
   \begin{equation}
B_3=\left(
\begin{matrix}
b_{11}&b_{12}&b_{13}&b_{14} \cr
 b_{21}&b_{22}&b_{23}&b_{24} \cr
 - (b_{13} b_{22}-b_{12} b_{23})  - b_{41} &0&0&-b_{44} \cr
 b_{41}&0&0& b_{44}
 \end{matrix}
\right),
\end{equation}
where, since $B_3 B_2^{-1} \in A^t  $, then, $ B_3$  is equivalent to $ B _2$. So, we obtained the Lie bialgebra
 $(A_{4,1},(II\oplus R).i)$ such that commutation relations for $(II\oplus R).i$
 are as follows:
 \begin{equation}
[{\tilde X}^1,{\tilde X}^2]= {\tilde X}^3+{\tilde X}^4.
\end{equation}

 Similarly, we use the above method for the classification of the four dimensional real  Lie bialgebras of symplectic type
 \footnote{Note that we consider only Lie bialgebras for  which there are symplectic structures on the Lie algebra ${\bf g}$
  and their duals $\tilde{\bf g}$, according to table 1 }.
  The results of such calculations are give in table 2 (Appendix A).\\


\section  {\large {\bf Four dimensional coboundary real  Lie bialgebras of symplectic type}}

In this section, we determine number of  four dimensional real
Lie bialgebras of symplectic type of  table 2 which are coboundary.
 For this purpose, we must find $r=r^{ij}X_i \otimes X_j
\in {\bf g}\otimes{\bf g}$ such that the  cocommutator of symplectic
Lie bialgebras can be written as (\ref{BA1}). Using (\ref{BA1}),
(\ref{BA2}) and (\ref{BA3}), we have \cite{RHR}
\begin{equation}\label{BA4}
{\tilde{\cal Y}}_i = {{\cal X}_i}^{t} r +r {\cal X}_i .
\end{equation}
  Now using  the above relations, one can find the
r-matrix of the Lie bialgebras of table 2 (if there exists).  One can also
perform this task for the dual Lie bialgebras $(\tilde{\bf
g}, {\bf g})$  using the following equations (\ref{BA4})
\begin{equation}
{\cal Y}^i = ({\tilde{\cal X}}^i)^{t} \tilde{r} +\tilde{r} {\tilde{\cal X}}^i .
\end{equation}
 The results are summarized in tables 3 and 4.
Notice that, we also determine the Schouten brackets of these r-matrices and consider the  classical  r-matrices, satisfying the
CYB equation. In table 3 of Appendix B, we have listed the triangular Lie
bialgebras $({\bf g},{\bf {\tilde{g}}})$ with triangular duals $({\bf
{\tilde{g}}},{\bf g})$. Since such structures can be specified (up to
automorphism) by pairs of r-matrices, then it is natural to call
them symplectic  bi-r-matrix Lie bialgebras \cite{RHR}. Furthermore,
we provide complete list of four dimensional triangular
Lie bialgebras of symplectic type in the table 4 (Appendix B).\\

\section {\large {\bf Calculation of  Poisson and symplectic  structures on four dimensional real Poisson-Lie groups }}
\smallskip

 \subsection {Calculation of  Poisson and symplectic  structures by Sklyanin
 bracket}

For the  Lie bialgebras one can obtain the corresponding Poisson-Lie
groups by means of
 Sklyanin bracket provided by a given   skew-symmetric
r-matrix $r=r^{ij}X_i \otimes X_j$ \cite{VC} as follows:
\begin{equation}\label{BA7}
\lbrace f_1\;,\;f_2\rbrace =\sum _{i,j} r^{ij}((X^L_i f_1)(X^L_j f_2)-(X^R_i f_1)(X^R_j f_2)),\;\;\;\;\;\;\forall f_1,f_2\in C^\infty (G),
\end{equation}
where $  ( X^{L }_i)$ and $(X^{R }_i)$ are left and right invariant vector fields  on the Poisson-Lie group $G$.
For calculation of the left and right invariant vector fields on the
Lie group $G$, it is enough to determine the left and right
invariant one forms. For $\forall g\in G$, we have
\begin{equation}
dg\;g^{-1}= R^i\;{X_i}\;\;\;\;\;\;(dg\;g^{-1})^{i}=R^i=R^i_{\;j}~ dx^j,
\end{equation}
\begin{equation}
g^{-1} dg =L^i\;X_i\;\;\;\;\;\;(g^{-1}\;dg)^{i}=L^i=L^i_{\;j} ~dx^j,
\end{equation}
where $x^i$ are coordinates on the Lie group. Now, from $\delta ^i_j=< X^R_j,R^i >$ and $ \delta ^i_j=<X^L_j,L^i>$ together with $X^R_j=X^{R \;l}_j \partial _l$ and $X^L_j=X^{L\;l}_j \partial _l$ , we have
\begin{equation}\label{BC1}
X_j^{R\;l}= (R^{-t})_j^{\;l},\;\;\;\;\;X_j^{L\;l}= (L^{-t})_j^{\;l}.
\end{equation}
To calculate the above matrices, we assume the following
parameterization of the  four dimensional real Lie group $G$:
\begin{equation}
g = e^{x_1 X_1}e^{x_2 X_2} e^{x_3 X_3} e^{x_4 X_4}.
\end{equation}
Then,  for left and right invariant Lie algebra valued one forms, we
have:
\begin{equation}
dg g^{-1}=dx_1 X_1+dx_2 e^{x_1 X_1}X_2 e^{-x_1X_1}+dx_3  e^{x_1 X_1}(e^{x_2 X_2}X_3 e^{-x_2 X_2}) e^{-x_1X_1}\hspace*{2.5cm}
\end{equation}
 $\hspace*{3.5cm }+dx_4 e^{x_1 X_1}e^{x_2 X_2}( e^{x_3 X_3}X_4 e^{-x_3 X_3})e^{-x_2 X_2} e^{-x_1X_1},$

\begin{equation}
 g^{-1} dg=dx_1 e^{-x_4 X_4}e^{-x_3 X_3}( e^{-x_2 X_2}X_1 e^{x_2 X_2})e^{x_3 X_3} e^{x_4 X_4}+dx_2  e^{-x_4 X_4}(e^{-x_3 X_3}X_2 e^{x_3 X_3}) e^{x_4 X_4}
\end{equation}
 $\hspace*{3.5cm}+dx_3 e^{-x_4 X_4}X_3 e^{x_4 X_4}+dx_4 X_4,$\\
 such that, for this calculation one can use the following relation

\begin{equation}
(e^{-x_i X_i}X_j e^{x_i X_i}) = (e^{x_i {\cal X}_i})_j ^{\;k} X_k,
\end{equation}
in which  we have a summation over the index $k$ on the right hand side. In this way, we  obtained the left and right invariant vector fields as
given in table 5 (Appendix C). Now, by using these results one can calculate the
Poisson structures over the  Lie group $G$. For simplicity we can
rewrite the relation (\ref{BA7}) in the following matrix form:
\begin{equation}
\{f_1,f_2\}=(X^L_1 f_1 \;\;\; X^L_2 f_1\;\;\;X^L_3 f_1\;\;\;X^L_4 f_1)\; r\; (X^L_1 f_2 \;\;\; X^L_2 f_2\;\;\;X^L_3 f_2\;\;\;X^L_4 f_2)^t
\end{equation}
  $\hspace*{4.55cm}-(X^R_1 f_1 \;\;\; X^R_2 f_1\;\;\;X^R_3 f_1\;\;\;X^R_4 f_1)\; r \;(X^R_1 f_2 \;\;\; X^R_2 f_2\;\;\;X^R_3 f_2\;\;\;X^R_4
  f_2)^t.$\\
Using the tables 5 and r-matrices presented in tables 3 and 4, one
can calculate  Poisson brackets on the Poisson-Lie groups $G$ and
${\tilde G}$. The  results are listed in table 6 (Appendix D).

\subsection  {\bf  Calculation of Poisson  and symplectic structures on four dimensional real
 Poisson-Lie group  by Manin
 triple and map
$\pi (g)$ between the Lie subalgebras $\bf g $ and $ \tilde{\bf g}$.
}
Some of the Lie bialgebras in table 2 do not have r-matrix, such that, for
obtaining the corresponding Poisson-Lie group one    can use the
following relation:
\begin{equation}
g^{-1} X_i g=a(g)_i^{\;j} X_j,\;\;\;\;\;\;g^{-1}{\tilde{X}}^i g
=b(g)^{ij} X_j + d(g)^i_{\;j} {\tilde{X}}^j,
\end{equation}
\begin{equation}
\pi (g) = b(g) a^{-1}(g),
\end{equation}
where $\pi (g)$ is the algebraic Poisson structure, such that, the
Poisson structure   on the Lie group $G$ can be obtained as follows:
\begin{equation}
P_{G} (g) = (- b(g) a(g)^{-1})^{ij} X_i^{R}\wedge  X_j^{R}.
\end{equation}
Therefore, using (\ref{BC1}),  we have the following relation for the Poisson
structures
\begin{equation}\label{BC2}
P^{kl}= (- b(g) a(g)^{-1})^{ij} X_i^{R\;k}  X_j^{R\;l}.
\end{equation}
Now, using the right invariant vector fields of table 5, and
calculating the adjoint matrices $a(g)$  and $b(g)$, one can
obtain the Poisson brackets on the non-coboundary Lie bialgebras
{\footnote{Note that using the relation (\ref{BC2}) for cobondary Lie
bialgebras one can obtain the same results which are obtained from Sklyanin
bracket (\ref{BA7})}}. The results are given in table 7 (Appendix D).

\subsection  {\bf  Compatible symplectic structure on four dimensional real Poisson-Lie groups  }
Among Poisson structures given in tables 6 and 7, one can consider
those with invertible matrices $P^{ij}$. In
this way, one arrives at the compatible symplectic structure on the
 four dimensional real Poisson-Lie groups. Such results are given in
tables 8 and 9.\\

{\small {\bf Table 8}}:{ \small Related Lie bialgebras with
compatible bi-symplectic\footnote{By means of bi-symplectic
structure is that there are symplectic structure on the
Lie groups $G$ and  $\tilde{G}$ related to Lie bialgebras $({\bf g},\tilde{\bf g})$ and $(\tilde{\bf g},{\bf g})$ respectively.} structure on four dimensional real Poisson-Lie groups}\\
    \begin{tabular}{|l| l| l |l| l|   }
    \hline\hline
{\footnotesize $({\bf g},\tilde{\bf g})$ }& {\footnotesize $({\bf g},\tilde{\bf g})$ }
&{\footnotesize $({\bf g},\tilde{\bf g})$ }&{\footnotesize $({\bf g},\tilde{\bf g})$ }&{\footnotesize $({\bf g},\tilde{\bf g})$ }\\

 \hline

{\footnotesize  $(A_{4,7},A_{4,7}.i)$ }& {\footnotesize $(A_{4,7},A_{4,9}^{\frac{-1}{2}}.i)$ }
&{\footnotesize $(A_{4,9}^0,A_{4,9}^0.iv)$ } &{\footnotesize $(A_{4,9}^0,A_{4,9}^0.v)$}&{\footnotesize  $(A_{4,9}^{\frac{-1}{2}},A_{4,9}^{\frac{-1}{2}}.ii)$} \\
\hline
{\footnotesize  $(A_{4,9}^{\frac{-1}{2}},A_{4,9}^{\frac{-1}{2}}.iii)$  }& {\footnotesize  $(A_{4,9}^{\frac{-1}{2}},A_{4,9}^{\frac{-1}{2}}.iv)$ }&
{\footnotesize$(A_{4,9}^1,A_{4,9}^{1}.i)$  } &{\footnotesize $(A_{4,9}^{b},A_{4,9}^{b}.i)$}&{\footnotesize $(A_{4,9}^{\frac{-1}{2}},A_{4,9}^1.ii)$ } \\
\hline
{\footnotesize $(A_{4,11}^b,A_{4,11}^b.i)$  }& {\footnotesize $(A_{4,12},A_{4,12}.i)$}&
{\footnotesize $(A_{4,12},A_{4,12}.ii)$ } &{\footnotesize $(A_{4,12},A_{4,12}.iii)$ }&{\footnotesize $(A_2\oplus A_2,(A_2\oplus A_2).iv)$} \\
\hline
{\footnotesize $(A_2\oplus A_2,(A_2\oplus A_2).v)$  }& {\footnotesize $(A_2\oplus A_2,(A_2\oplus A_2).vi)$}&
{\footnotesize $(A_2\oplus A_2,(A_2\oplus A_2).vii)$} &{\footnotesize  }&{\footnotesize } \\
\hline\hline
\end{tabular}\\

{\small {\bf Table 9}}:{ \small
 Related Lie bialgebras with compatible symplectic structure on four dimensional real Poisson-Lie groups}\\
    \begin{tabular}{|l| l| l |l|    }
    \hline\hline
{\footnotesize $({\bf g}\;,\;\tilde{\bf g})$ }& {\footnotesize $({\bf g}\;,\;\tilde{\bf g})$ }
&{\footnotesize $({\bf g} \;,\;\tilde{\bf g})$ }&{\footnotesize $({\bf g}\;,\;\tilde{\bf g})$ }\\
 \hline
{\footnotesize $(A_{4,1}\;,\; A_{4,9}^0.i)$ }& {\footnotesize  $(A_{4,1}\;,\; A_{4,9}^0.ii)$ }
&{\footnotesize  $(A_{4,1}\;,\; A_{4,9}^0.iii)$ }&{\footnotesize $(A_{4,3}\;,\;(A_2 \oplus A_2).i)$ }\\
\hline
{\footnotesize $(A_{4,3}\;,\;(A_2 \oplus A_2).iii)$}& {\footnotesize $((II\oplus R).iv \;,\; A_{4,7})$ }
&{\footnotesize $((II\oplus R).ii\;,\; A_{4,7})$ }&{\footnotesize $((II\oplus R).x \;,\; A_{4,9}^0)$}\\
 \hline
 {\footnotesize $((II\oplus R).iv \;,\; A_{4,9}^0)$ }& {\footnotesize $((II\oplus R).ii \;,\; A_{4,9}^0)$ }
&{\footnotesize $((II\oplus R).viii \;,\; A_{4,9}^0)$ }&{\footnotesize $((II\oplus R).xi \;,\; A_{4,9}^0)$ }\\
 \hline
 {\footnotesize  $((II\oplus R).ii \;,\; A_{4,9}^{-\frac{1}{2} })$  }& {\footnotesize $((II\oplus R).ii \;,\; A_{4,9}^1)$ }
&{\footnotesize $((II\oplus R).iv \;,\; A_{4,9}^b)$ }&{\footnotesize $((II\oplus R).ii \;,\; A_{4,11}^b)$ }\\
 \hline
 {\footnotesize $((II\oplus R).xii \;,\; A_{4,12})$ }& {\footnotesize $((II\oplus R).xiii \;,\; A_2 \oplus A_2)$ }
&{\footnotesize $((III\oplus R).i \;,\; A_2 \oplus A_2)$ }&{\footnotesize  $((III\oplus R).iii \;,\; A_2 \oplus A_2)$ }\\
 \hline
 {\footnotesize  $((III\oplus R).v \;,\; A_2 \oplus A_2)$ }& {\footnotesize  $((VI_0 \oplus R).vi \;,\; A_2 \oplus A_2)$ }
&{\footnotesize $((VI_0 \oplus R).vii \;,\; A_2 \oplus A_2)$}&{\footnotesize $((VI_0 \oplus R).viii \;,\; A_2 \oplus A_2)$ }\\
 \hline
 {\footnotesize $( (VI_0\oplus R).v\;,\; A_{4,9}^0)$ }& {\footnotesize $( (VII_0\oplus R).ii\;,\;A_{4,9}^0)$ }
&{\footnotesize $((II\oplus R).iv \;,\; A_{4,9}^{-\frac{1}{2} })$ }&{\footnotesize $((II\oplus R).xii \;,\; A_{4,12})$ }\\
  \hline
{\footnotesize  $((III\oplus R).iv \;,\; A_2 \oplus A_2)$ }&
{\footnotesize $( (VI_0\oplus R).iv\;,\;A_{4,9}^0)$ }&{\footnotesize $(A_{4,3}\;,\;(A_2 \oplus A_2).ii)$ }&
{\footnotesize  }\\
\hline\hline
\end{tabular}

\section {\large {\bf Physical application; example of  integrable systems by use of symplectic  Lie bialgebra }}

Here, we consider some integrable systems obtained by using the
symplectic Lie bialgebras. In these examples, we consider the Lie group $G$
related to the Lie bialgebra $({\bf g}, {\bf {\tilde{g}})}$ as a phase space and its dual Lie group ${\tilde{G}}$ as a
symmetry group of the system. For this propose we use the formalism
mentioned
in \cite{JAF} for calculation of  two integrable systems with some symmetry groups ${\tilde{G}}$. Here the Lie groups
 $\bf{A_{4,9}^{-\frac{1}{2}}} $ and $\bf{A_{4,9}^1.ii}$  play the role of phase  space and symmetry group, respectively.\\

{\bf Example 1)}  The Lie group $\bf{A_{4,9}^{-\frac{1}{2}}} $ as a phase space and
 $\bf{A_{4,9}^1.ii}$
as symmetry group :\\
 For this example, the Darboux coordinates have
the following forms \cite{JAF}:
 \begin{equation}
 y_1=x_2,\;\;\;\; y_2=\frac{e^{-x_4}(x_1+x_2 x_3)}{x_2},\;\;\;\;\; y_3=\frac{e^{-\frac{x_4}{2}}(-x_1+e^{\frac{x_4}{2}} x_1-x_2 x_3)}{2x_2^2},\;\;\;\; y_4=e^{\frac{x_4}{2}},
 \end{equation}
 such that in this Darboux coordinates the symplectic structure of  $\bf{A_{4,9}^{-\frac{1}{2}}} $  (according table 6)

 \begin{equation}\label{BA111}
 \lbrace x_1,x_2\rbrace =-2 x_2^2,\;\;\lbrace x_1,x_3\rbrace =-x_1+x_2 x_3,\;\;\lbrace x_1,x_4\rbrace =2x_2,\;\;\lbrace x_2,x_3\rbrace =-2 x_2,\;\; \lbrace x_3,x_4\rbrace =-2+2 e^{ x_4/2},
 \end{equation}
can be simplified as follows:
\begin{equation}\label{BA1112}
 \lbrace y_1,y_3\rbrace =1 , \hspace{0.5cm} \lbrace
y_2,y_4\rbrace =1 .
\end{equation}
 In this way, we have the following forms for the dynamical functions $Q_i$ according to \cite{JAF}
\begin{equation}
Q_1=-y_4=-e^{\frac{x_4}{2}} ,\;\;\;\;\;\;\;\;\;\;\;Q_2=-\frac{y_3}{2}=-\frac{e^{-\frac{x_4}{2}}(-x_1+e^{\frac{x_4}{2}} x_1-x_2 x_3)}{4x_2^2},
\end{equation}

$$Q_3=2y_1 y_3+y_2 y_4=\frac{x_1}{x_2},\;\;\;Q_4=-y_2 y_3=\frac{e^{-3x_4/2} x_1^2}{2 x_2^3}-\frac{e^{-x_4}x_1^2}{2x_2^3}+\frac{e^{-3 x_4/2}x_1 x_3}{x_2^2}-\frac{e^{-x_4}x_1 x_3}{2 x_2^2}+\frac{e^{-3x_4/2}x_3^2}{2x_2}, $$
such that, they satisfy the following Poisson brackets by use of
(\ref{BA111})  or (\ref{BA1112}) as follows:
\begin{equation}
 \lbrace Q_1,Q_3\rbrace =-Q_1\;,\; \lbrace Q_1,Q_4\rbrace =2Q_2
 \;,\; \lbrace Q_2,Q_3\rbrace =-2Q_2\;,\; \lbrace Q_3,Q_4\rbrace
 =Q_4,
\end{equation}
i.e., a Poisson bracket$ \lbrace Q_i,Q_j \rbrace =f_{ij}^k Q_k  $, where, $f_{ij}^k$ are  the structure constants of the symmetry Lie
algebra  ${A_{4,9}^{1}.ii}$. The invariants of the above system are $(Q_1,Q_2)$ or $(Q_2,Q_4)$, such that one can consider one of these $Q_i$ as Hamiltonian of the integrable systems.\\

{\bf Example 2)} In this example the role of phase space and  symmetry group of example 1 are interchanged, i.e., the Lie group $\bf{A_{4,9}^{1}.ii}$ plays the role of  phase space and $\bf{A_{4,9}^{-\frac{1}{2}}}$ plays the role of  as symmetry group :\\
For this example, the Darboux coordinates have the following forms \cite{JAF}:
\begin{equation}
 y_1=x_1,\;\;\;\; y_2=-\frac{2e^{x_3}x_1 x_4+ x_3}{x_1},\;\;\;\;\; y_3=\frac{-x_2+e^{-x_3}x_2+2x_1 x_4}{x_1^2},
 \;\;\;\; y_4=e^{-x_3},
\end{equation}
such that, we have the following  forms for the symplectic structure
(according table 6):
\begin{equation}\label{BA112}
 \lbrace x_1,x_2\rbrace=-x_1^2 \; ,\;\; \lbrace x_1,x_4\rbrace=\frac{1}{2} e^{-x_3}x_1 \; ,\;\;
 \lbrace x_2,x_3\rbrace=x_1 \; ,\;\;\lbrace x_2,x_4\rbrace=e^{-x_3}x_2 \; ,\;\;
 \lbrace x_3,x_4\rbrace=\frac{-1}{2}(1-e^{-x_3})\; ,\;\;
\end{equation}
also, we have the following forms for the dynamical functions $Q_i$ \cite{JAF}
\begin{equation}
Q_1=-y_3=-\frac{-x_2+e^{-x_3}x_2+2x_1 x_4}{x_1^2} ,\;\;\;\;\;\;\;\;Q_2=- y_4=-e^{-x_3} ,
\end{equation}

$Q_3=-y_2 y_3=(\frac{2e^{x_3}x_1 x_4+ x_3}{x_1})(\frac{-x_2+e^{-x_3}x_2+2x_1 x_4}{x_1^2}),\;\;\;\;\;
Q_4=-\frac{1}{2} y_1 y_3-y_2 y_4=\frac{x_2}{2x_1}+\frac{e^{-x_3}x_2}{x_1}+x_4,$\\
with them satisfying  the following Poisson brackets according to
(\ref{BA112}) as follows:
\begin{equation}
 \lbrace Q_1,Q_4\rbrace =\frac{1}{2}Q_1\;,\; \lbrace Q_2,Q_3\rbrace =Q_1
 \;,\; \lbrace Q_2,Q_4\rbrace =Q_2\;,\; \lbrace Q_3,Q_4\rbrace
 =-\frac{1}{2}Q_3,
\end{equation}
i.e., a Poisson bracket$ \lbrace Q_i,Q_j \rbrace =f_{ij}^k Q_k  $, where, $f_{ij}^k$ are  the structure constants of the symmetry Lie
algebra  $A_{4,9}^{-\frac{1}{2}}$.
Where the invariants of the system are $(Q_1,Q_2)$ or $(Q_1,Q_3)$,
such that one can consider one of these $Q_i$ as Hamiltonian of the integrable
systems.

\section {\large {\bf Conclusion}}
We classify all four dimensional real  Lie bialgebras of symplectic type and
obtain the classical r-matrices, Poisson and symplectic
structures on all of the corresponding four dimensional Poisson-Lie
groups. We also give two examples as the physical application, such that for
these integrable systems the Poisson-Lie group $G$ plays the role of
phase space and its dual Lie group ${\tilde{G}}$ plays the  role of
symmetry group of the system. Calculation of  all such systems and
also systems for which  the role of  $G$ and ${\tilde{G}}$ are
replaced and  the relation between such systems is under
investigation.\\

{\bf Acknowledgments}\\
We would like to thank M. Akbari-Moghanjoughi for carefully reading the manuscript and 
M. Sephid for their useful comments.

\newpage
\smallskip
{\bf Appendix A:}\\

{\small {\bf Table 2}}: {\small
Four dimensional real   Lie bialgebras of symplectic type.}\\
\begin{tabular}{l l l l  p{0.15mm} }
    \hline\hline
{\footnotesize ${\bf g}$ }& {\footnotesize $\tilde{\bf g}$}
&{\footnotesize Non-zero commutation relations of
$\tilde{\bf g}$}& {\footnotesize Comments} \\ \hline

\vspace{2mm}

{\footnotesize $A_{4,1} $}&{\footnotesize $4A_1 $}&
\\

\vspace{1mm}

&{\footnotesize$A_{4,1}.i $}&{\footnotesize$[{\tilde
X}^1,{\tilde X}^2]= {\tilde X}^3,\;\;\;[{\tilde X}^2,{\tilde
X}^3]= {\tilde X}^4$} &\\

\vspace{1mm}

&{\footnotesize$A_{4,1}.ii $}&{\footnotesize$[{\tilde
X}^1,{\tilde X}^2]= {\tilde X}^4,\;\;\;[{\tilde X}^1,{\tilde
X}^4]= {\tilde X}^3$} &\\

\vspace{1mm}

&{\footnotesize$A_{4,1}.iii $}&{\footnotesize$[{\tilde
X}^1,{\tilde X}^2]= {\tilde X}^3,\;\;\;[{\tilde X}^1,{\tilde
X}^3]= {\tilde X}^4$} &\\

\vspace{1mm}

&{\footnotesize$A^0_{4,9}.i $}&{\footnotesize$[{\tilde
X}^1,{\tilde X}^2]= {\tilde X}^1+{\tilde X}^2,\;\;\;[{\tilde X}^1,{\tilde
X}^4]= {\tilde X}^4,\;\;\;[{\tilde X}^2,{\tilde
X}^3]=q {\tilde X}^4,\;\;\;[{\tilde X}^2,{\tilde
X}^4]=- {\tilde X}^4$} &{\footnotesize $q\in
{\Re-\{0\}}$}\\

\vspace{1mm}

&{\footnotesize$A^0_{4,9}.ii $}&{\footnotesize$[{\tilde
X}^1,{\tilde X}^2]= {\tilde X}^1,\;\;\;[{\tilde X}^1,{\tilde
X}^3]= {\tilde X}^4,\;\;\;[{\tilde X}^2,{\tilde
X}^4]=- {\tilde X}^4$} &\\

\vspace{1mm}

&{\footnotesize$A^0_{4,9}.iii $}&{\footnotesize$[{\tilde
X}^1,{\tilde X}^2]= {\tilde X}^2,\;\;\;[{\tilde X}^1,{\tilde
X}^4]= {\tilde X}^4,\;\;\;[{\tilde X}^2,{\tilde
X}^3]= {\tilde X}^4$} &\\
\hline\hline
\end{tabular}
\newpage

{\small {\bf Table 2}}: {\small
 (Continued.)}\\
    \begin{tabular}{l l l l  p{0.15mm} }
    \hline\hline
{\footnotesize ${\bf g}$ }& {\footnotesize $\tilde{\bf g}$}
&{\footnotesize Non-zero commutation relations of
$\tilde{\bf g}$}& {\footnotesize Comments} \\ \hline
\vspace{1mm}

\vspace{1mm}

{\footnotesize $A^{-1,b}_{4,5} $}&{\footnotesize $4A_1 $}&\\

\vspace{1mm}

&{\footnotesize$A^{-1,b}_{4,5}.i $}&{\footnotesize$[{\tilde
X}^1,{\tilde X}^3]={\tilde X}^1,\;\;\;[{\tilde X}^2,{\tilde
X}^3]=- {\tilde X}^2,\;\;\;[{\tilde X}^3,{\tilde
X}^4]= b{\tilde X}^4$} &\\

\vspace{1mm}

&{\footnotesize$(II\oplus R).iv $}&{\footnotesize$[{\tilde
X}^1,{\tilde X}^3]={\tilde X}^4,$} &\\

\vspace{1mm}

&{\footnotesize$(II\oplus R).ii $}&{\footnotesize$[{\tilde
X}^1,{\tilde X}^2]={\tilde X}^4,$} &\\

\vspace{1mm}

&{\footnotesize$(II\oplus R).viii $}&{\footnotesize$[{\tilde
X}^1,{\tilde X}^2]={\tilde X}^4,\;\;\;[{\tilde X}^1,{\tilde
X}^3]={\tilde X}^4$} &\\

\vspace{1mm}

&{\footnotesize$(II\oplus R).iii $}&{\footnotesize$[{\tilde
X}^2,{\tilde X}^3]={\tilde X}^4,$} &\\

\vspace{1mm}

&{\footnotesize$(II\oplus R).v $}&{\footnotesize$[{\tilde
X}^1,{\tilde X}^2]={\tilde X}^4,\;\;\;[{\tilde X}^2,{\tilde
X}^3]={\tilde X}^4$} &\\
\hline\hline
\end{tabular}
\newpage
{\small {\bf Table 2}}: {\small
 (Continued.)}\\
    \begin{tabular}{l l l l  p{0.15mm} }
    \hline\hline
{\footnotesize ${\bf g}$ }& {\footnotesize $\tilde{\bf g}$}
&{\footnotesize Non-zero commutation relations of
$\tilde{\bf g}$}& {\footnotesize Comments} \\ \hline

\vspace{1mm}

&{\footnotesize$A^{-\frac{1}{2}}_{4,9}.iii $}&{\footnotesize$[{\tilde
X}^1,{\tilde X}^2]=-2{\tilde X}^2,\;\;\;[{\tilde X}^1,{\tilde
X}^3]=4{\tilde X}^3,\;\;\;[{\tilde X}^1,{\tilde
X}^4]= 2{\tilde X}^4,\;\;\;[{\tilde X}^2,{\tilde
X}^3]= {\tilde X}^4$} &\\

\vspace{1mm}

&{\footnotesize$A^{-\frac{1}{2}}_{4,9}.iv $}&{\footnotesize$[{\tilde
X}^1,{\tilde X}^2]=\frac{1}{2}{\tilde X}^2,\;\;\;[{\tilde X}^1,{\tilde
X}^4]={\tilde X}^3,\;\;\;[{\tilde X}^2,{\tilde
X}^3]= \frac{1}{2}{\tilde X}^3,\;\;\;[{\tilde X}^2,{\tilde
X}^4]= {\tilde X}^4$} &\\

\vspace{1mm}

&{\footnotesize$A^{1}_{4,9}.ii $}&{\footnotesize$[{\tilde
X}^1,{\tilde X}^3]=-{\tilde X}^1,\;\;\;[{\tilde X}^1,{\tilde
X}^4]=2{\tilde X}^2,\;\;\;[{\tilde X}^2,{\tilde
X}^3]= -2{\tilde X}^2,\;\;\;[{\tilde X}^3,{\tilde
X}^4]= {\tilde X}^4$} &\\

\vspace{1mm}

&{\footnotesize$(II\oplus R).iv$}&{\footnotesize$[{\tilde X}^1,{\tilde
X}^3]={\tilde X}^4,$} &\\

\vspace{1mm}

&{\footnotesize$(II\oplus R).ii $}&{\footnotesize$[{\tilde X}^1,{\tilde
X}^2]={\tilde X}^4,$} &\\

\vspace{1mm}


{\footnotesize $A^{1}_{4,9} $}&{\footnotesize $4A_1 $}&
\\

\vspace{1mm}

&{\footnotesize$A^{1}_{4,9}.i $}&{\footnotesize$[{\tilde
X}^1,{\tilde X}^2]=-\frac{1}{4}{\tilde X}^2,\;\;\;[{\tilde X}^1,{\tilde
X}^3]=-\frac{1}{4}{\tilde X}^3,\;\;\;[{\tilde X}^1,{\tilde
X}^4]= -\frac{1}{2}{\tilde X}^4,\;\;\;[{\tilde X}^2,{\tilde
X}^3]= -{\tilde X}^4$} &\\
\hline\hline
\end{tabular}
\newpage

{\small {\bf Table 2}}: {\small
 (Continued.)}\\
    \begin{tabular}{l l l l  p{0.15mm} }
    \hline\hline
{\footnotesize ${\bf g}$ }& {\footnotesize $\tilde{\bf g}$}
&{\footnotesize Non-zero commutation relations of
$\tilde{\bf g}$}& {\footnotesize Comments} \\ \hline

{\footnotesize $VII_0\oplus R $}&{\footnotesize $4R $}&\\

\vspace{1mm}

&{\footnotesize $(VII_0\oplus R).iv$}&{\footnotesize$[{\tilde
X}^1,{\tilde X}^4]={\tilde X}^2,\;\;\;[{\tilde X}^2,{\tilde
X}^4]=-{\tilde X}^1,$} &\\

\vspace{1mm}

&{\footnotesize $(II\oplus R).xiv$}&{\footnotesize$[{\tilde
X}^1,{\tilde X}^4]={\tilde X}^3,$} &\\

\vspace{1mm}

&{\footnotesize $(II\oplus R).xii$}&{\footnotesize$[{\tilde
X}^1,{\tilde X}^2]={\tilde X}^3,$} &\\

\vspace{1mm}

&{\footnotesize $(II\oplus R).i$}&{\footnotesize$[{\tilde
X}^1,{\tilde X}^2]={\tilde X}^3+{\tilde X}^4,$} &\\

\vspace{1mm}


{\footnotesize $III\oplus R $}&{\footnotesize $4A_1 $}&
\\

\vspace{1mm}

&{\footnotesize $(III\oplus R).xiii$}&{\footnotesize$[{\tilde
X}^1,{\tilde X}^2]={\tilde X}^1,\;\;\;[{\tilde X}^1,{\tilde
X}^3]={\tilde X}^1,\;\;\;[{\tilde X}^2,{\tilde
X}^3]={\tilde X}^1,$} &\\
\hline\hline
\end{tabular}
\smallskip

\vspace*{3mm}
{\bf Appendix B:}\\

{ {\bf Table 3}}: ~{\small Four dimensional coboundary and
bi-r-matrix  Lie bialgebras of symplectic type.\smallskip\\
\smallskip
\begin{tabular}{l l l  l l  p{1mm} }
    \hline\hline
\smallskip
{\footnotesize $({\bf g}, \tilde{\bf g})$ }&{\footnotesize  $r$}
&{\footnotesize $[[r , r]]$}&{\footnotesize ${\tilde
r}$}&{\footnotesize $[[\tilde r , \tilde r]]$}\\
\hline
\smallskip

\vspace{2mm}

{\footnotesize $(A_{4,1},A_{4,1}.iii)$}&
{\footnotesize  $a X_1 \otimes X_1+c( X_1 \wedge X_3-X_2 \otimes X_2)+$} &{\footnotesize $0$} &{\footnotesize $ d({\tilde X}^2 \otimes {\tilde
X}^4 +{\tilde X}^4 \otimes {\tilde
X}^2-{\tilde X}^3 \otimes {\tilde
X}^3)$} &{\footnotesize $0$} \\
\vspace{2mm}
&{\footnotesize  $b  X_1 \wedge X_2+X_1 \wedge X_4-X_2 \wedge X_3$} & &{\footnotesize $+{\tilde X}^2 \wedge {\tilde
X}^3+e {\tilde X}^3 \wedge {\tilde X}^4+{\tilde
X}^4 \wedge {\tilde X}^1$} & \\
\vspace{2mm}
&{\footnotesize  } & &{\footnotesize $+f {\tilde X}^4 \otimes {\tilde X}^4$} & \\

{\footnotesize $(A^{-1}_{4,2},A^{-1}_{4,2}.i)$}&
{\footnotesize  $a X_1 \otimes X_2+X_1 \wedge X_3+b X_2 \otimes X_1-$} &{\footnotesize $0$} &{\footnotesize $-{\tilde X}^1 \wedge {\tilde X}^3+{\tilde X}^2 \wedge {\tilde X}^4 +c {\tilde X}^3 \otimes {\tilde X}^4$} &{\footnotesize $0$} \\
\vspace{2mm}
&{\footnotesize $X_2 \wedge X_4$ } & &{\footnotesize $+d{\tilde X}^4 \otimes {\tilde X}^3$} & \\

\vspace{2mm}

{\footnotesize $(A_{4,3},A_{4,3}.ii)$}&
{\footnotesize  $ -X_1 \wedge X_4+a  X_2 \otimes X_2+b X_2 \wedge X_3$} &{\footnotesize $0$} &{\footnotesize ${\tilde X}^1 \wedge {\tilde X}^4+c{\tilde X}^2 \wedge {\tilde X}^3 +d{\tilde X}^3 \otimes {\tilde X}^3$} &{\footnotesize $0$} \\

\vspace{2mm}

{\footnotesize $(A^{-1,-1}_{4,5},A^{-1,-1}_{4,5}.i)$}&
{\footnotesize  $a X_1 \otimes X_2+b  X_1 \otimes X_3+c X_2 \otimes X_1+$} &{\footnotesize $0$} &{\footnotesize $e{\tilde X}^1 \otimes {\tilde X}^3-{\tilde X}^2 \wedge {\tilde X}^4 +f{\tilde X}^3 \otimes {\tilde X}^1+$} &{\footnotesize $0$} \\
\hline\hline
\end{tabular}
{ {\bf Table 3}}: ~{\small (Continued.)}\smallskip\\
\smallskip
\begin{tabular}{l l l  l l  p{1mm} }
    \hline\hline
\smallskip
{\footnotesize $({\bf g}, \tilde{\bf g})$ }&{\footnotesize  $r$}
&{\footnotesize $[[r , r]]$}&{\footnotesize ${\tilde
r}$}&{\footnotesize $[[\tilde r , \tilde r]]$}\\
\hline
\smallskip

{\footnotesize $(A^{-\frac{1}{2}}_{4,9},A^{1}_{4,9}.ii)$}&
{\footnotesize  $a X_1 \wedge X_3 +2 X_3 \wedge X_4$} &{\footnotesize $0$} &{\footnotesize $-\frac{1}{2}{\tilde X}^3 \wedge {\tilde X}^4$} &{\footnotesize $0$} \\

\vspace{2mm}

{\footnotesize $(A^{1}_{4,9},A^{1}_{4,9}.i)$}&
{\footnotesize  $\frac{1}{4}  X_1 \wedge X_4 +\frac{1}{2} X_2 \wedge X_3$} &{\footnotesize $0$} &{\footnotesize $-4{\tilde X}^1 \wedge {\tilde X}^4-2 {\tilde X}^2 \wedge {\tilde X}^3$} &{\footnotesize $0$} \\

\vspace{2mm}

{\footnotesize $(A^{b}_{4,9},A^{b}_{4,9}.i)$}&
{\footnotesize  $- X_1 \wedge X_4 -(1+b)X_2 \wedge X_3$} &{\footnotesize $0$} &{\footnotesize ${\tilde X}^1 \wedge {\tilde X}^4+\frac{1}{1+b} {\tilde X}^2 \wedge {\tilde X}^3$} &{\footnotesize $0$} \\

\vspace{2mm}

{\footnotesize $(A^{b}_{4,11},A^{b}_{4,11}.i)$}&
{\footnotesize  $-\frac{1}{2b} X_1 \wedge X_4 -X_2 \wedge X_3$} &{\footnotesize $0$} &{\footnotesize $2b{\tilde X}^1 \wedge {\tilde X}^4+ {\tilde X}^2 \wedge {\tilde X}^3$} &{\footnotesize $0$} \\

\vspace{2mm}

{\footnotesize $(A_{4,12},A_{4,12}.ii)$}&
{\footnotesize  $-\frac{q}{2} X_1 \wedge X_2 -X_1 \wedge X_4+X_2 \wedge X_3$} &{\footnotesize $0$} &{\footnotesize ${\tilde X}^1 \wedge {\tilde X}^4- {\tilde X}^2 \wedge {\tilde X}^3-\frac{q}{2}{\tilde X}^3\wedge {\tilde X}^4$} &{\footnotesize $0$} \\

{\footnotesize $(A_2 \oplus A_2,(A_2 \oplus A_2).v)$}&
{\footnotesize  $-q X_1 \wedge X_3$} &{\footnotesize $0$} &{\footnotesize $\frac{1}{q}{\tilde X}^1\wedge {\tilde X}^3$} &{\footnotesize $0$} \\
\hline\hline
\end{tabular}

\vspace{3.5mm}

{\bf Table 4}:~{\small Four dimensional coboundary  Lie bialgebras of symplectic type.}\\
 \begin{tabular}{l l l l p{0.5mm} }
  \hline\hline

{\footnotesize $({\bf g}, \tilde{\bf g})$ }& r
& [[r,r]]\\
\hline
\smallskip
\vspace{1mm}
{\footnotesize $( A_{4,1},4A_1)$}&{\footnotesize $ a X _1 \otimes X_1+ b X_1\wedge X_2+c (X_1 \wedge X_3- X_2 \otimes X_2)$}&{\footnotesize $0$}\\

\vspace{2mm}

{\footnotesize $( A_{4,1},(II\oplus R).i)$}& {\footnotesize
$ a X _1 \otimes X_1+ b X_1\wedge X_2+c X_1 \otimes X_3+X_1\wedge X_4-(1-c) X_2 \otimes X_2 +(2+c) X_3 \otimes X_1$}&
{\footnotesize $0$}\\

\vspace{2mm}

{\footnotesize $( A_{4,1},(II\oplus R).ii)$}& {\footnotesize
$ a X _1 \otimes X_1+ b X_1\wedge X_2+c X_1 \otimes X_3-(1+c) X_2 \otimes X_2 +(2+c) X_3 \otimes X_1$}&
{\footnotesize $0$}\\

\vspace{2mm}

{\footnotesize $( A_{4,1},(II\oplus R).iv)$}& {\footnotesize
$ a X _1 \otimes X_1+ b X_1\wedge X_2+c( X_1 \wedge X_3-X_2 \otimes X_2)- X_2 \wedge X_3  $}&
{\footnotesize $0$}\\

\vspace{2mm}

{\footnotesize $( A^{-1}_{4,2},4A_1)$}& {\footnotesize
$ a X_1 \otimes X_2+b X_2 \otimes X_1$}&
{\footnotesize $0$}\\

\vspace{2mm}

{\footnotesize $( A^{-1}_{4,2},A^{-1,-1}_{4,5}.i)$}& {\footnotesize
$ a X_1 \otimes X_2+b X_2 \otimes X_1 -X_2 \wedge X_4$}&
{\footnotesize $0$}\\

\vspace{2mm}

{\footnotesize $( A^{-1}_{4,2},(II\oplus R).iii)$}& {\footnotesize
$  a X_1 \otimes X_2+b X_2 \otimes X_1 -\frac{1}{2} X_2 \wedge X_3$}&
{\footnotesize $0$}\\
\hline\hline
\end{tabular}

{\bf Table 4}:~{\small (Continued.)}\\
 \begin{tabular}{l l l l p{0.5mm} }
  \hline\hline
{\footnotesize $({\bf g}, \tilde{\bf g})$ }& r
& [[r,r]]\\
\hline
\smallskip
\vspace{2mm}

{\footnotesize $( A^{a,-1}_{4,5},4A_1)$}& {\footnotesize
$  c X_1 \otimes X_3+d X_3 \otimes X_1$}&
{\footnotesize $0$}\\

\vspace{2mm}

{\footnotesize $( A^{a,-1}_{4,5},(II\oplus R).iii)$}& {\footnotesize
$  c X_1 \otimes X_3+d X_3 \otimes X_1 + \frac{1}{1-a}  X_2\wedge X_3$}&
{\footnotesize $0$}\\

\vspace{2mm}

{\footnotesize $( A^{a,-1}_{4,5},(II\oplus R).v)$}& {\footnotesize
$  -\frac{1}{1+a} X_1 \wedge X_2+c X_1 \otimes X_3+d X_3 \otimes X_1 - \frac{1}{a-1}  X_2\wedge X_3$}&
{\footnotesize $0$}\\

\vspace{2mm}

{\footnotesize $( A^{a,-a}_{4,5},4A_1)$}& {\footnotesize
$c X_2 \otimes X_3+d X_3 \otimes X_2$}&
{\footnotesize $0$}\\

 \vspace{2mm}

{\footnotesize $( A^{a,-a}_{4,5},(II\oplus R).iv)$}& {\footnotesize
$  -\frac{1}{1-a} X_1 \wedge X_3+c X_2 \otimes X_3+d X_3 \otimes X_2$}&
{\footnotesize $0$}\\

 \vspace{2mm}

{\footnotesize $( A^{a,-a}_{4,5},(II\oplus R).viii)$}& {\footnotesize
$  -\frac{1}{1+a} X_1 \wedge X_2-\frac{1}{1-a} X_1\wedge X_3+c X_2 \otimes X_3+d X_3 \otimes X_2$}&
{\footnotesize $0$}\\

 \vspace{2mm}

{\footnotesize $( A^{a,0}_{4,6},4A_1)$}& {\footnotesize
$  c( X_2 \otimes X_2+ X_3 \otimes X_3)+d X_2 \wedge X_3$}&
{\footnotesize $0$}\\
\hline\hline
\end{tabular}

{\bf Appendix C:}\\
{\small {\bf Table 5}}: {\small
Left and right invariant vector fields over four dimensional real Lie groups of symplectic type.}\\
    \begin{tabular}{|l| l| l|  }
    \hline\hline
{\footnotesize ${\bf g}$ }&{\footnotesize $\left(\begin{array}{c}X^L_1\\ X^L_2\\ X^L_3\\ X^L_4\end{array} \right)$ } &{\footnotesize $\left(\begin{array}{c}X^R_1\\ X^R_2\\ X^R_3\\ X^R_4\end{array} \right)$ } \\ \hline

{\footnotesize ${ A_{4,1}}$ }&{\footnotesize $\left(\begin{array}{c} \partial _1\\ \partial _2 -x_4 \partial _1\\ \partial _3-x_4\partial _2+\frac{x_4^2}{2}\partial _1 \\ \partial _4\end{array} \right)$ } &{\footnotesize $\left(\begin{array}{c} \partial _1\\ \partial _2 \\ \partial _3\\ \partial _4 -x_2 \partial _1-x_3 \partial _2\end{array} \right)$ } \\ \hline

{\footnotesize ${ A_{4,1}.iii}$ }&{\footnotesize $\left(\begin{array}{c} \partial _1 - x_3\partial _4-x_2 \partial _3\\ \partial _2 \\ \partial _3  \\ \partial _4\end{array} \right)$ } &{\footnotesize $\left(\begin{array}{c} \partial _1\\ \partial _2 -x_1 \partial _3+\frac{x_1 ^2}{2} \partial _4 \\ \partial _3 - x_1 \partial _4 \\ \partial _4 \end{array} \right)$ } \\ \hline

{\footnotesize ${ A_{4,2}^{-1}}$ }&{\footnotesize $\left(\begin{array}{c} e^{x_4}\partial _1 \\ e^{-x_4}\partial _2 \\ e^{-x_4}(\partial _3 -x_4 \partial _2) \\ \partial _4\end{array} \right)$ } &{\footnotesize $\left(\begin{array}{c} \partial _1\\ \partial _2 \\ \partial _3  \\  x_1 \partial _1 -(x_2+x_3)\partial _2
-x_3\partial _3+ \partial _4 \end{array} \right)$ } \\ \hline

{\footnotesize ${ A_{4,2}^{-1}.i}$ }&{\footnotesize $\left(\begin{array}{c} e^{x_2}\partial _1 + x_2 \partial _4 \\ \partial _2 -x_4 \partial _4+x_3 \partial _3 \\ \partial _3 \\ \partial _4\end{array} \right)$ } &{\footnotesize $\left(\begin{array}{c} \partial _1\\ \partial _2  + x_1 \partial _1 +x_1 e^{-x_2}\partial\partial _4\\ e^{x_2} \partial _3  \\ e^{-x_2} \partial _4\end{array} \right)$ } \\
\hline

\end{tabular}

\newpage
{\small {\bf Table 5}}: {\small
(Continued.)}\\
    \begin{tabular}{|l| l| l|  }
    \hline\hline

{\footnotesize ${ A_{4,7}.i}$ }&{\footnotesize $\left(\begin{array}{c}\partial _1 - \frac{x_2}{2} \partial _2 +\frac{x_2-x_3}{2}\partial _3-\frac{x_2^2+2x_4}{2} \partial _4  \\ \partial _2-2x_3 \partial _4  \\ \partial _3 \\ \partial _4\end{array} \right)$ } &{\footnotesize $\left(\begin{array}{c} \partial _1\\ e^{\frac{-x_1}{2}}(\partial _2+\frac{x_1}{2}\partial _3-x_1x_2\partial _4) \\ e^{\frac{-x_1}{2}}(\partial _3-2x_2\partial _4) \\\ e^{-x_1}\partial _4\end{array} \right)$ } \\ \hline

{\footnotesize ${ A_{4,9}^{0}}$ }&{\footnotesize $\left(\begin{array}{c}e^{-x_4}\partial _1  \\ e^{- x_4}(\partial _2-x_3 \partial _1) \\ \partial _3 \\ \partial _4\end{array} \right)$ } &{\footnotesize $\left(\begin{array}{c} \partial _1\\ \partial _2  \\ \partial _3  -x_2 \partial _1\\ -x_1\partial _1 - x_2 \partial _2+\partial _4\end{array} \right)$ } \\ \hline

{\footnotesize ${ A_{4,9}^{0}.iv}$ }&{\footnotesize $\left(\begin{array}{c}\partial _1  - x_3 \partial _3-(x_2+x_4) \partial _4  \\  \partial _2
-x_3 \partial _4  \\ \partial _3 \\ \partial _4\end{array} \right)$ } &{\footnotesize $\left(\begin{array}{c} \partial _1\\ \partial _2+(e^{-x_1}
-1)\partial _4  \\ e^{- x_1} (\partial _3 -x_2 \partial _4) \\ e^{- x_1}\partial _4\end{array} \right)$ } \\ \hline

{\footnotesize ${ A_{4,9}^{-1/2}}$ }&{\footnotesize $\left(\begin{array}{c}e^{-\frac{x_4}{2}}\partial _1  \\ e^{- x_4}(\partial _2-x_3 \partial _1) \\ e^{\frac{x_4}{2}}\partial _3 \\ \partial _4\end{array} \right)$ } &{\footnotesize $\left(\begin{array}{c} \partial _1\\ \partial _2  \\ \partial _3  -x_2 \partial _1\\-\frac{x_1}{2}\partial _1 - x_2 \partial _2+\frac{x_3}{2}\partial _3+\partial _4\end{array} \right)$ } \\
\hline

\end{tabular}

\newpage
{\small {\bf Table 5 }}: {\small
(Continued.)}\\
    \begin{tabular}{|l| l| l|  }
    \hline\hline

{\footnotesize ${VII_0\oplus  R}$ }&{\footnotesize $\left(\begin{array}{c}cos( x_3)\partial _1+sin( x_3)\partial _2  \\ cos( x_3)\partial _2-sin( x_3)\partial _1\\ \partial _3 \\ \partial _4\end{array} \right)$ } &{\footnotesize $\left(\begin{array}{c} \partial _1\\ \partial _2 \\ -x_2\partial _1 +x_1\partial _2 +\partial _3 \\ \partial _4\end{array} \right)$ } \\ \hline

{\footnotesize ${(VII_0\oplus  R).iv}$ }&{\footnotesize
$\left(\begin{array}{c}cos(x_4)\partial _1-sin(x_4)\partial _2  \\
 cos(x_4)\partial _2+sin(x_4)\partial _1 \\ \partial _3 \\
  \partial _4\end{array} \right)$ } &{\footnotesize $\left(\begin{array}{c}
   \partial _1\\ \partial _2  \\ \partial _3 \\
   x_2 \partial _1-x_1 \partial _2 +\partial _4\end{array} \right)$ } \\ \hline

{\footnotesize ${(VII_0\oplus  R).ii}$ }&{\footnotesize
$\left(\begin{array}{c}\partial _1-x_4\partial _2 +x_2 \partial _4 \\
\partial _2 \\ \partial _3 \\
  \partial _4\end{array} \right)$ } &{\footnotesize $\left(\begin{array}{c}
   \partial _1\\ cos(x_1)\partial _2 +sin(x_1)\partial _4 \\ \partial _3 \\
  cos(x_1)\partial _4 -sin(x_1)\partial _2 \end{array} \right)$ } \\ \hline

{\footnotesize ${A_{4,11}^b}$ }&{\footnotesize
$\left(\begin{array}{c} e^{-2b x_4}\partial _1  \\e^{-b x_4}((-x_3
\partial _1+\partial _2)cos(x_4)+sin( x_4)\partial _3)\\ e^{-b
x_4}((x_3 \partial _1-\partial _2)sin(x_4)+cos (x_4)\partial _3)
\\\partial _4 \end{array} \right)$ } &{\footnotesize
$\left(\begin{array}{c} \partial _1\\ \partial _2 \\-x_2
\partial _1+ \partial _3 \\ \frac{x_3^3-x_2^2-4b x_1}{2} \partial
_1- (x_3+b x_2 )\partial _2+(x_2-b x_3)\partial _3+\partial
_4\end{array} \right ) $ } \\
\hline

\end{tabular}

{\bf Appendix D:}\\

{\small {\bf Table 6}}: {\small Poisson brackets over Poisson-Lie
group $G$ (using Sklyanin
bracket ) .}\\
    \begin{tabular}{l l l  p{0.15mm} }
    \hline\hline
{\footnotesize ${\bf g}$ }& {\footnotesize $\tilde{\bf g}$}
&{\footnotesize Non-zero Poisson brackets relations } \\ \hline

{\footnotesize $A_{4,1} $}&{\footnotesize $4A_1 $}\\

\vspace{0.7mm}

&{\footnotesize$A_{4,1}.iii $}&{\footnotesize$ \lbrace x_1,x_2\rbrace=x_3-\frac{x_4^4}{2},\;\;\;\lbrace x_1,x_3\rbrace=x_4$} \\

\vspace{0.7mm}

&{\footnotesize$(II\oplus R).i $}&{\footnotesize$\lbrace x_1,x_2\rbrace=x_3+x_4$} \\

\vspace{0.7mm}

&{\footnotesize$(II\oplus R).ii $}&{\footnotesize$ \lbrace x_1,x_2\rbrace=x_4$} \\

\vspace{0.7mm}

&{\footnotesize$(II \oplus R).iv $}&{\footnotesize$\lbrace x_1,x_2\rbrace=-\frac{x_4^2}{2},\;\;\;\lbrace x_1,x_3\rbrace=x_4 $} \\


{\footnotesize $A^{-1}_{4,2} $}&{\footnotesize $4A_1 $}\\

\vspace{0.7mm}

&{\footnotesize$A^{-1}_{4,2}.i $}&{\footnotesize$\lbrace x_1,x_2\rbrace=-x_1-x_4,\;\;\;\lbrace x_2,x_3\rbrace=-x_3,\;\;\;\lbrace x_2,x_4\rbrace=1- e^{-x_4}$} \\
\hline\hline
\end{tabular}

\newpage
{\small {\bf Table 6}}: {\small
 (Continued.)}\\
    \begin{tabular}{l  l l  p{0.15mm} }
    \hline\hline
{\footnotesize ${\bf g}$ }& {\footnotesize $\tilde{\bf g}$}
&{\footnotesize Non-zero Poisson brackets relations } \\ \hline

\vspace{0.7mm}

{\footnotesize $A^{-\frac{1}{2}}_{4,9} $}&{\footnotesize $4A_1 $}&\\

\vspace{0.7mm}

&{\footnotesize$A^{-\frac{1}{2}}_{4,9}.iii $}&{\footnotesize$\lbrace x_1,x_2\rbrace=-2x_2,\;\;\lbrace x_1,x_3\rbrace= 2x_3+2e^{\frac{-x_4}{2}}x_3,\;\;\lbrace x_1,x_4\rbrace=4-4e^{\frac{-x_4}{2}},\;\;\lbrace x_2,x_3\rbrace=2-2e^{\frac{-x_4}{2}}$} \\

\vspace{0.7mm}

&{\footnotesize$A^{-\frac{1}{2}}_{4,9}.iv $}&{\footnotesize$\lbrace x_1,x_2\rbrace=\frac{x_1}{2},\;\;\lbrace x_1,x_4\rbrace=x_3e^{-x_4},\;\;\lbrace x_2,x_3\rbrace=\frac{x_3}{2},\;\;\lbrace x_2,x_4\rbrace=1-e^{-x_4} $} \\

\vspace{0.7mm}

&{\footnotesize$A^{1}_{4,9}.ii $}&{\footnotesize$\lbrace x_1,x_2\rbrace=-2x_2^2, \lbrace x_1,x_3\rbrace=x_2 x_3-x_1, \lbrace x_1,x_4\rbrace=2x_2, \lbrace x_2,x_3\rbrace=-2x_2, \lbrace x_3,x_4\rbrace=2e^{\frac{x_4}{2}}-2 $} \\

\vspace{0.7mm}

&{\footnotesize$(II\oplus R).ii $}&{\footnotesize$ \lbrace x_1,x_2\rbrace=\frac{2}{3}(1-e^{-\frac{3x_4}{2}})$} \\

\vspace{0.7mm}


{\footnotesize $A^{1}_{4,9} $}&{\footnotesize $4A_1 $}&
\\

\vspace{0.7mm}

&{\footnotesize$A^{1}_{4,9}.i $}&{\footnotesize$ \lbrace x_1,x_2\rbrace=\frac{-x_2}{4}, \lbrace x_1,x_3\rbrace=\frac{x_3}{4}(1-2 e^{-2x_4}), \lbrace x_4,x_1\rbrace=\frac{1}{4}(1-e^{-2x_4}), \lbrace x_3,x_2\rbrace=\frac{1}{2}(1-e^{-2x_4})$} \\
\hline\hline
\end{tabular}
\newpage
{\small {\bf Table 6}}: {\small
 (Continued.)}\\
    \begin{tabular}{l  l l  p{0.15mm} }
    \hline\hline
{\footnotesize ${\bf g}$ }& {\footnotesize $\tilde{\bf g}$}
&{\footnotesize Non-zero Poisson brackets relations } \\
\hline
\vspace{0.7mm}

{\footnotesize$A_{4,9}^0.iv$}&{\footnotesize$A_{4,9}^0$}&{\footnotesize$\lbrace x_1,x_4\rbrace= 1- e^{-x_1} \;,\; \lbrace x_2,x_3\rbrace=1- e^{-x_1} \;,\; \lbrace x_2,x_4\rbrace=e^{-x_1} x_2 \;,\; \lbrace x_3,x_4\rbrace= \frac{1}{2}+\frac{e^{-2 x_1}}{2} -  e^{-x_1}$} \\

 \vspace{1mm}
{\footnotesize$(VI_0\oplus R).v$}&{\footnotesize$A_{4,9}^0$}&{\footnotesize$\lbrace x_1,x_4\rbrace= -1+ e^{x_1} \;,\; \lbrace x_2,x_3\rbrace=1- e^{-x_1} \;,\; \lbrace x_2,x_4\rbrace= x_2 $} \\

\vspace{1mm}

{\footnotesize$A_{4,9}^{-\frac{1}{2}}.iii$}&{\footnotesize$A_{4,9}^{-\frac{1}{2}}$}&{\footnotesize$\lbrace x_1,x_4\rbrace=\frac{1}{4}( 1- e^{-2 x_1} )\;,\; \lbrace x_2,x_3\rbrace=\frac{1}{2}(1- e^{-2x_1} )\;,\; \lbrace x_2,x_4\rbrace=\frac{x_2}{2}(1+e^{-2x_1}  )\;,\; \lbrace x_3,x_4\rbrace= -\frac{x_3}{2}$} \\

\vspace{1mm}

{\footnotesize$A_{4,9}^{-\frac{1}{2}}.iv$}&{\footnotesize$A_{4,9}^{-\frac{1}{2}}$}&{\footnotesize$\lbrace x_1,x_3\rbrace=-\frac{1}{2}e^{ \frac{-x_2}{2}} x_1^2 , \lbrace x_1,x_4\rbrace=\frac{1}{2}e^{-x_2} x_1 , \lbrace x_2,x_3\rbrace=e^{\frac{-x_2}{2}} x_1 , \lbrace x_2,x_4\rbrace= 1-e^{x_2}, \lbrace x_3,x_4\rbrace= \frac{-x_3}{2}$} \\

\vspace{1mm}

{\footnotesize$A_{4,9}^{1}.ii$}&{\footnotesize$A_{4,9}^{-\frac{1}{2}}$}&{\footnotesize$\lbrace x_1,x_2\rbrace=- x_1^2 , \lbrace x_1,x_4\rbrace=\frac{1}{2}e^{-x_3}x_1 , \lbrace x_2,x_3\rbrace=x_1, \lbrace x_2,x_4\rbrace= e^{-x_3} x_2 ,  \lbrace x_3,x_4\rbrace= \frac{-1}{2}(1-e^{-x_3} )$} \\

\vspace{1mm}

{\footnotesize$A_{4,9}^{1}.i$}&{\footnotesize$A_{4,9}^{1}$}&{\footnotesize$\lbrace x_1,x_4\rbrace=-4(1-e^{\frac{x_1}{2}}) \;,\; \lbrace x_2,x_3\rbrace=-2(1-e^{\frac{x_1}{2}})\;,\; \lbrace x_2,x_4\rbrace=-x_2(1-2e^{\frac{x_1}{2}}) \;,\; \lbrace x_3,x_4\rbrace= x_3$} \\
\hline\hline
\end{tabular}

\vspace{5mm}
{\small {\bf Table 7}}: {\small
Poisson brackets over Poisson-Lie group $G$ (using $\pi (g)$).}\\
    \begin{tabular}{l l l  p{0.15mm} }
    \hline\hline
{\footnotesize ${\bf g}$ }& {\footnotesize $\tilde{\bf g}$}
&{\footnotesize Non-zero Poisson brackets relations } \\
 \hline

{\footnotesize $A_{4,1} $}&{\footnotesize$A_{4,1}.i $}&{\footnotesize$\lbrace x_1,x_2\rbrace=x_3+\frac{x_4^3}{6},\;\;\;\lbrace x_1,x_3\rbrace=-\frac{x_4^2}{2},\;\;\;\lbrace x_2,x_3\rbrace=x_4$} \\

\vspace{1mm}

&{\footnotesize$A_{4,1}.ii $}&{\footnotesize$ \lbrace x_1,x_2\rbrace=-\frac{x_3^2}{2}+x_4,\;\;\;\lbrace x_1,x_4\rbrace=x_3$} &\\

\vspace{1mm}

&{\footnotesize$A^0_{4,9}.i $}&{\footnotesize$\lbrace x_1,x_2\rbrace=x_1+x_2+\frac{q x_4^3}{6},\;\lbrace x_1,x_3\rbrace=-\frac{q x_4^2}{2},\;\lbrace x_1,x_4\rbrace=x_4+ \frac{x_4^2}{2},\;\lbrace x_2,x_3\rbrace=q x_4,\;\lbrace x_2,x_4\rbrace=-x_4$} \\

\vspace{1mm}

&{\footnotesize$A^0_{4,9}.ii $}&{\footnotesize$ \lbrace x_1,x_2\rbrace=x_1-\frac{x_4^2}{2},\;\;\lbrace x_1,x_3\rbrace=x_4,\;\;
\lbrace x_1,x_4\rbrace=\frac{x_4^2}{2},\;\;\lbrace x_2,x_4\rbrace=-x_4$} \\

\vspace{1mm}

&{\footnotesize$A^0_{4,9}.iii $}&{\footnotesize$\lbrace x_1,x_2\rbrace=x_2+\frac{x_4^3}{6},\;\;\lbrace x_1,x_3\rbrace=
-\frac{x_4^2}{2},\;\;\lbrace x_1,x_4\rbrace=x_4,\;\;\lbrace x_2,x_3\rbrace=x_4$} \\

\vspace{1mm}

&{\footnotesize$(II\oplus R).iii $}&{\footnotesize$ \lbrace x_1,x_2\rbrace=\frac{x_4^3}{6},\;\;\;\lbrace x_1,x_3\rbrace=-\frac{x_4^2}{2},\;\;\;\lbrace x_2,x_3\rbrace=x_4$} \\
\vspace{1mm}

&{\footnotesize$(VI_0\oplus R).i $}&{\footnotesize$\lbrace x_1,x_2\rbrace=-\frac{x_3^2+x_4^2}{2},\;\;\;\lbrace x_1,x_3\rbrace= x_4,\lbrace x_1,x_4\rbrace=x_3$} \\
\hline\hline
\end{tabular}

\newpage

{\small {\bf Table 7}}: {\small
 (Continued.)}\\
    \begin{tabular}{l l l  p{0.15mm} }
    \hline\hline
{\footnotesize ${\bf g}$ }& {\footnotesize $\tilde{\bf g}$}
&{\footnotesize Non-zero Poisson brackets relations}\\
\hline
\vspace{1mm}

&{\footnotesize$(II\oplus R).v $}&{\footnotesize$\lbrace x_1,x_2\rbrace=1-e^{-x_4},\;\;\;\lbrace x_2,x_3\rbrace=x_4 $} \\

\vspace{1mm}

&{\footnotesize$(II\oplus R).vi $}&{\footnotesize$ \lbrace x_1,x_2\rbrace=-1+e^{-x_4}(1+x_4),\;\;\;\lbrace x_1,x_3\rbrace=1-e^{-x_4},\;\;\;\lbrace x_2,x_3\rbrace=x_4$} \\

\vspace{1mm}

&{\footnotesize$(III\oplus R).ii $}&{\footnotesize$\lbrace x_2,x_3\rbrace=x_3$} \\

\vspace{1mm}

&{\footnotesize$(VI_0\oplus R).iii $}&{\footnotesize$ \lbrace x_1,x_2\rbrace=x_1,\;\;\;\lbrace x_2,x_3\rbrace=x_3$} \\


\vspace{1mm}

{\footnotesize $A^{-1,-1}_{4,5} $}&{\footnotesize$(II\oplus R).ii $}&{\footnotesize$\lbrace x_1,x_2\rbrace=x_4$} \\

\vspace{1mm}

{\footnotesize $A^{-1,b}_{4,5} $}&{\footnotesize$(II\oplus R).ii $}&{\footnotesize$ \lbrace x_1,x_2\rbrace=x_4$} \\

\vspace{1mm}

&{\footnotesize$(II\oplus R).viii $}&{\footnotesize$ \lbrace x_1,x_2\rbrace=x_4,\;\;\;\lbrace x_1,x_3\rbrace=\frac{1-e^{-(1+b)x_4}}{1+b}$} \\
\hline\hline
\end{tabular}

\newpage

{\small {\bf Table 7}}: {\small
 (Continued.)}\\
    \begin{tabular}{l l l  p{0.15mm} }
    \hline\hline
{\footnotesize ${\bf g}$ }& {\footnotesize $\tilde{\bf g}$}
&{\footnotesize Non-zero Poisson brackets relations}\\
\hline
\vspace{1mm}

&{\footnotesize $(II\oplus R).xvii$}&{\footnotesize$ \lbrace x_1,x_2\rbrace=-x_3,\;\;\;\lbrace x_1,x_4\rbrace=-x_3-\frac{1}{2} x_3^2,\;\;\;\lbrace x_2,x_4\rbrace=x_3$} \\

\vspace{1mm}

&{\footnotesize $(II\oplus R).xviii$}&{\footnotesize$ \lbrace x_1,x_2\rbrace=-x_3$} \\

\vspace{1mm}

&{\footnotesize $(II\oplus R).ii$}&{\footnotesize$ \lbrace x_1,x_2\rbrace=x_4 $} \\

\vspace{0.5mm}
{\footnotesize $(II\oplus R).iv$}&{\footnotesize $A_{4,7}$}&{\footnotesize $ \lbrace x_1,x_4\rbrace=2x_1\;,\;\lbrace x_2,x_3\rbrace=x_1
\;,\;\lbrace x_2,x_4\rbrace=-\frac{x_1^2}{2} + x_2\;,\;\lbrace x_3,x_4\rbrace=x_2+x_3$} \\

\vspace{0.5mm}
{\footnotesize $(II\oplus R).ii$}&{\footnotesize  $A_{4,7}$}&{\footnotesize $ \lbrace x_1,x_4\rbrace=2x_1\;,\;\lbrace x_2,x_3\rbrace=x_1
\;,\;\lbrace x_2,x_4\rbrace= x_2\;,\;\lbrace x_3,x_4\rbrace=\frac{x_1^2}{2}+x_2+x_3$} \\

\vspace{0.5mm}
{\footnotesize $(II\oplus R).x$}&{\footnotesize $A_{4,9}^0$}&{\footnotesize $ \lbrace x_1,x_4\rbrace=x_1\;,\;\lbrace x_2,x_3\rbrace=x_1
\;,\;\lbrace x_2,x_4\rbrace= -\frac{x_1^2}{2} +x_2$} \\

\vspace{0.5mm}
{\footnotesize $(II\oplus R).viii$}&{\footnotesize $A_{4,9}^0$}&{\footnotesize $ \lbrace x_1,x_4\rbrace=x_1\;,\;\lbrace x_2,x_3\rbrace=x_1
\;,\;\lbrace x_2,x_4\rbrace=-\frac{x_1^2}{2} +x_2\;,\;\lbrace x_3,x_4\rbrace=\frac{x_1^2}{2}$} \\
\hline\hline
\end{tabular}

\end{document}